\documentclass[11pt]{article}
\usepackage[utf8]{inputenc}
\usepackage{hyperref}
\usepackage{graphicx}
\usepackage{dcolumn}
\usepackage{bm}
\usepackage{longtable}
\usepackage{amsmath}
\RequirePackage{wasysym} \RequirePackage{setspace}\textheight=650pt
\pdfoutput=1
\begin{document}
\title{ {\bf Electric quadrupole transitions of the Bohr Hamiltonian with Manning-Rosen potential }}
\author{M. Chabab$^{a}$, A. El Batoul$^{b}$, A. Lahbas$^{c}$, M. Oulne$^{d*}$ \\
\\ {\small High Energy Physics and Astrophysics Laboratory, Faculty of Science Semlalia, }\\
{\small Cadi Ayyad University, P.O.B. 2390, Marrakesh,Morocco} \\
{\small $^{a}$ mchabab@uca.ma} \\
{\small $^{b}$ elbatoul.abdelwahed@edu.uca.ma} \\
{\small $^{c}$ alaaeddine.lahbas@edu.uca.ma} \\
{\small $^{d*}$ corresponding author  oulne@uca.ma}\\
}

\maketitle

\begin{abstract}
Analytical expressions of the  wave functions are derived for a Bohr Hamiltonian with the Manning–Rosen potential in the cases of $\gamma$-unstable nuclei and axially symmetric prolate deformed ones with $\gamma \approx 0$. By exploiting the  results we have obtained in a recent work on the same theme  Ref. \cite{CLO15MR}, we have calculated the $B(E2)$ transition rates for 34 $\gamma$-unstable and 38 rotational nuclei and compared to experimental data, revealing a qualitative agreement with the experiment and  phase transitions within the ground state band and showing also that the Manning–Rosen potential is more appropriate for such calculations than other potentials.

\end{abstract}
\section{Introduction}
The advent of critical point symmetries in nuclear structure has motivated the search for adequate approaches with different potentials within the Bohr-Mottelson model \cite{bohr,boh75} for describing nuclei which are close to or away from these symmetries. Among these symmetries, one can cite for example the critical point symmetry E(5) \cite{E5} which allows describing the second order phase transition between vibrational and $\gamma$-unstable nuclei and the X(5) \cite{X5} critical point symmetry corresponding to the first order phase transition between vibrational and axially symmetric prolate deformed nuclei. In this context, different potential models have been used such as  infinite square well potential \cite{E5,budaca15},  Morse potential \cite{bozt8,Inci11}, Kratzer potential \cite{Fortunato05,bonat13}, Davidson potential \cite{bonat7,bonat11}, Sextic potential\cite{levai2004,levai2010,buganu15sextic,buganu12,buganu2011,Raduta:2013db,Buganu:2013}, Quartic potential \cite{budaca14}, Woods-Saxon \cite{capak15}, Hulthen potential \cite{CLO15H}...etc. Recently, we have used the Manning-Rosen potential to calculate energy spectra and the ground state bandhead $R_{4/2} = E(4^+_{g})/E(2^+_{g})$ ratios, as well as those of the $\beta$ and $\gamma$ bandheads, normalized to the $2^+_{g}$ for nuclei with a mass number $100<A<250$ \cite{CLO15MR}. The corresponding energy formula has been obtained in closed analytical form  by means of the asymptotic iteration method \cite{AIM03,AIM05,AIM07} as well as the corresponding normalized wave functions. This method has proved to be a useful tool when dealing
with physical problems involving Schr\"odinger-type equations \cite{CO10,chabab2012exact,CHABAB:2012qd,chabab2015exact,chabab2016closed}. In order to extend that work to include transition probabilities calculations, in this paper we  exploit the already obtained normalized wave functions \cite{CLO15MR} to evaluate transition rates in the different bands for several $\gamma$-unstable and axially symmetric prolate deformed nuclei. The obtained numerical results are in a good agreement with the experimental data and are generally better than those obtained by Bonatsos et al. using Morse \cite{Inci11}, Kratzer \cite{bonat13} and Davidson potentials \cite{bonat11} for the same nuclei.

The present paper is organized as follows. In Section \ref{sec1} the normalized wave functions are constructed and the electric quadrupole transition rates are calculated for $\gamma$-unstable and axially symmetric prolate deformed nuclei, while Section \ref{sec2} is devoted to the numerical calculations of the $B(E2)$ transition probabilities along with their comparisons with the experimental data and with results from other models. Finally, Section \ref{sec3} contains the conclusion.

\section{B(E2) Transition rates \label{sec1}}
\subsection{The $\gamma$-unstable nuclei}
The original collective Bohr Hamiltonian is given by \cite{bohr}
\begin{equation}
H=-\frac{\hbar ^2}{2B}\left[ \frac{1}{\beta^4}\frac{\partial}{\partial\beta} {\beta^4}\frac{\partial}{\partial\beta}+ \frac{1}{\beta^2\sin3\gamma}\frac{1}{\partial\gamma}\sin3\gamma\frac{\partial}{\partial\gamma}-
  \frac{1}{4\beta^2}\sum_{k}\frac{Q_{k}^{2}}{\sin^2(\gamma-\frac{2}{3}\pi k)} \right]+V(\beta,\gamma)
  \label{1}
\end{equation}
where $\beta$ and $\gamma$ are the usual collective coordinates,  $Q_k (k=1, 2, 3) $ are the components of angular momentum in the intrinsic frame, and $B$ is the mass parameter.

In the case of $\gamma$-unstable potentials, we assume that the potential $V(\beta,\gamma)$ depends only on the coordinate $\beta$, i.e. $V(\beta,\gamma)=U(\beta)$. This kind of potentials are appropriate to describe the nuclei which can depart from axial symmetry without any energy cost.
The total wave function can be constructed as  \cite{wilet}
\begin{equation}
\Psi(\beta,\gamma,\theta_i)=\xi(\beta)\Phi(\gamma,\theta_i)
 \label{2}
\end{equation}
where $\theta_i(i=1,2,3)$ are the Euler angles. By introducing the reduced energy $\epsilon=2BE/\hbar^2$ and reduced potentials $u=2BV/\hbar^2$, the corresponding Schr\"{o}dinger equation can be separately written as
 \begin{equation}
  \left[ -\frac{1}{\beta^4}\frac{\partial}{\partial\beta} {\beta^4}\frac{\partial}{\partial\beta}+u(\beta)+\frac{\tau(\tau+3)}{\beta^2}\right]\xi(\beta)=\epsilon \xi(\beta)  \label{3}
\end{equation}
\begin{equation}
 \left[- \frac{1}{\sin3\gamma}\frac{\partial}{\partial\gamma}\sin3\gamma\frac{\partial}{\partial\gamma}+
  \frac{1}{4}\sum_{k}\frac{Q_{k}^{2}}{\sin^2(\gamma-\frac{2}{3}\pi k)}\right]\Phi(\gamma,\theta_i)=\tau(\tau+3)\Phi(\gamma,\theta_i)  \label{4}
\end{equation}
where $\tau$ is the seniority quantum number, characterizing the irreducible representation of SO(5) and taking the values $\tau=0, 1, 2, ...$ \cite{rakavy}. The last equation has been solved by B\'es \cite{bes} long time ago.

Concerning the radial part \eqref{3}, the Rosen-Manning potential $u(\beta)$  \cite{manningrosen}  is used
\begin{equation}
u(\beta)=- \left(1-\frac{e^{\beta_{0}\alpha}-1}{e^{\beta\alpha}-1}\right)^2\label{5}
\end{equation}
where $\alpha$ is a screening parameter characterizing the range of the potential and the parameter $\beta_0$ indicates the position of the minimum of the potential leading to the energy spectrum \cite{CLO15MR}
\begin{equation}
\epsilon _{n,\tau}=-\frac{1}{4}\left[\frac{\big(\alpha^2(\mu+n)^2-(A+2)^2\big)\big(\alpha^2(\mu+n)^2-A^2\big)}{\alpha^2(\mu+n)^2}\right]
\label{6}
\end{equation}
where $n$ is the principal quantum number, and
\begin{align}
\mu=\frac{1}{2}+\sqrt{\frac{9}{4}+\tau(\tau + 3)+\frac{A^2}{\alpha^2}}, \   A=e^{\alpha\beta_0}-1
\label{7}
\end{align}
To determine the radial wave function one needs the parametrization \cite{CLO15MR}
\begin{align}
\xi(\beta)&=\beta^{-2}\chi(\beta), && y=e^{-\alpha\beta}, &&  \nonumber \\ \chi(y)&=y^\nu(1-y)^\mu f(y), && \nu=\frac{\sqrt{1-\epsilon_{n,\tau}}}{\alpha},
\label{8}
\end{align}

leading to
\begin{align}
f''(y)=\frac{(1+2\mu+2\nu)y-(1+2\nu)}{y(1-y)}f'(y)
+\frac{\alpha^2(\nu+\mu)^2+\epsilon-e^{2\alpha\beta_0}}{\alpha^2y(1-y)}f(y)
\label{9}
\end{align}
The solution of this equation is found to be
 \begin{equation}
f(y)=N_{n,\tau \ 2 }F_1(-n,n+2\mu+2\nu;1+2\nu;y)  \label{10}
\end{equation}
where $_2F_1$ are hyper-geometrical functions and $N_{n,\tau}$ is a normalization constant.

Therefore, according to the relation between hyper-geometrical functions and the
generalized Jacobi polynomials \cite{szego} the radial wave function in the $\gamma$-unstable case can be written as,
 \begin{align}
\chi(z)=&N_{n,\tau}2^{-\nu-\mu}\frac{\Gamma(1+n)\Gamma(1+2\nu)}{\Gamma(1+n+2\nu)}(1-z)^{\nu}(1+z)^{\mu}P_{n}^{(2\nu,2\mu-1)}(z), \     z=1-2y  \label{11}
\end{align}
The normalization constant computed via  the normalization condition,
 \begin{align}
 \int_0^{\infty} \xi^2(\beta)\beta^4 d\beta&= \int_0^{\infty} \chi^2(\beta) d\beta \nonumber \\&=\frac{1}{\alpha}\int_{-1}^{1} \frac{1}{1-z} \chi^2(z) dz=1 & \label{12}
 \end{align}
Using the usual orthogonality relation of Jacobi polynomials (see Eqs. (7.391.1) and (7.391.5) of  \cite{gradshteyn}) we get
\begin{align}
N_{n,\tau}=\left(\frac{\mu+n}{2\alpha n!\nu(\nu+\mu+n)} \right)^{-1/2}  \left( \frac{\big(\Gamma(2\nu+1)\Gamma(1+n)\big)^2}{\Gamma(2\nu+1+n)} \frac{\Gamma(2\mu+n)}{\Gamma(2\nu+2\mu+n)}\right)^{-1/2}  \label{13}
\end{align}
Having the expression of the total wave function, one can easily compute the $B(E2)$ transition rates. In the general case the quadrupole operator is defined as \cite{wilet}
 \begin{align}
      T_{M}^{(E2)}=t\alpha_2&=t\beta\left[\mathcal{D}^{(2)}_{M,0}(\theta_i)\cos\gamma  +\frac{1}{\sqrt{2}}\Big( \mathcal{D}^{(2)}_{M,2}(\theta_i)
      +\mathcal{D}^{(2)}_{M,-2}(\theta_i) \Big)\sin\gamma \right]
  \label{14}
\end{align}
 where $\mathcal{D}(\theta_i)$ denotes the Wigner functions of Euler angles and $t$ is a scale factor.

The $B(E2)$ transition rates from an initial to a final state are given by \cite{edmonds}
      \begin{align}
B(E2;s_i,L_i  \rightarrow s_f,L_f)  =\frac{5}{16\pi} \frac{\mid \left<s_f,L_f\mid\mid T^{(E2)} \mid\mid s_i,L_i\right>\mid^2}{(2L_i+1)}\nonumber\\
    =\frac{2L_f+1}{2L_i+1}B(E2;s_f,L_f  \rightarrow s_i,L_i)
  \label{15}
\end{align}
where $s$ denotes quantum numbers other than the angular momentum $L$.
Then, the full symmetrized wavefunction is written from equation \eqref{2} as
\begin{equation}
\Psi(\beta,\gamma,\theta_i)=\beta^{-2}\chi_{n,\tau}(\beta)\Phi_{\tau}(\gamma,\theta_i)
 \label{16}
\end{equation}
The radial function $\chi(\beta)$ is given by Eq. \eqref{11}, while the angular functions $\Phi_{\tau}(\gamma,\theta_i)$ have the form \cite{bes}
\begin{equation}
\Phi_{\tau}(\gamma,\theta_i)=\frac{1}{4\pi}\sqrt{\frac{(2\tau+3)!!}{\tau!}}\left( \frac{\alpha_2}{\beta^2}\right)^{\tau}
 \label{17}
\end{equation}
where $\alpha_2$ is defined in Eq. \eqref{14}. From Eqs. \eqref{15} and \eqref{17} one obtains \cite{bonat04}
\begin{equation}
B(E2;L_{n,\tau}\rightarrow(L+2)_{n',\tau+1})=\frac{(\tau+1)(4\tau+5)}{(2\tau+5)(4\tau+1)}t^2I_{n',\tau+1;n,\tau}^2
\label{18}
\end{equation}
with
\begin{equation}
I_{n',\tau+1;n,\tau}= \int_0^{\infty} \beta \xi_{n',\tau+1}(\beta)\xi_{n,\tau}(\beta)\beta^4d\beta
\label{19}
\end{equation}
In the special case of the Manning-Rosen potential and by using the eigenfunctions shown in Eq. \eqref{11}, the integrals of Eq. \eqref{19} take the form
\begin{equation}
I_{n',\tau+1;n,\tau}=\frac{1}{\alpha^2} \int_{-1}^{1} \left(\frac{\ln{\frac{1-z}{2}}}{1-z} \right) \chi_{n',\tau+1}(z)\chi_{n,\tau}(z)   dz
\label{20}
\end{equation}
\subsection{The axially symmetric prolate deformed nuclei}
Exact separation of the variables $\beta$ and $\gamma$ may be achieved when the reduced potential is chosen as in Refs. \cite{bozt8,bonat13,bonat11,Raduta:2013db,wilet,buganu2015analytical} $v(\beta,\gamma)=u(\beta)+w(\gamma)/\beta^2$. For the $\gamma$-part, we use a harmonic oscillator potential \cite{X5,bonat11} $w(\gamma)=(3c)^2\gamma^2$. In the same context, we consider a wave function of the form \cite{X5}
\begin{equation}
\Psi(\beta,\gamma,\theta_i)=F_L(\beta)\eta_K(\gamma)\mathcal{D}_{M,K}^L(\theta_i) \label{21}
\end{equation}
$L$  is the the total angular momentum, where  $M$ and $K$ are the eigenvalues of the projections of angular momentum on the laboratory fixed $x$-axis and the body-fixed $x'$-axis respectively.

By solving the $\gamma$-vibrational part of the Schr\"{o}dinger equation following method of \cite{CLO15P}, the $\gamma$ angular wave functions can be written as \cite{CLO15P}
\begin{equation}
\eta_{n_{\gamma},K}=N_{n_{\gamma},K}\ \gamma^{|K/2|}\ e^{-\frac{3c\gamma^2}{2}}L_{\tilde n_{\gamma}}^{|K/2|}\Big(3c\gamma^2\Big)
\label{22}
\end{equation}
with $\tilde n_{\gamma}=\frac{n_{\gamma}-|K/2|}{2}$ where $n_{\gamma}$ is the quantum number related to $\gamma$ oscillations, while $L_{\tilde n_{\gamma}}^{K/2}$ represents the Laguerre polynomial and $N_{\gamma,K}$ the normalization constant, determined from the normalization condition
\begin{equation}
\int_{0}^{\pi/3}\eta^2_{n_{\gamma},K}(\gamma)|\sin3\gamma|d\gamma=1
\label{23}
\end{equation}
In the case of small $\gamma$ vibration $|\sin3\gamma| \simeq |3\gamma|$, then the integral Eq. (\ref{23}) is easily calculated by using Eq. (8.980) of ~\cite{gradshteyn}. \\
This leads to
\begin{equation}
N_{n_{\gamma},K}=\left[\frac{2}{3}(3c)^{1+|K/2|}\frac{\tilde n_{\gamma}!}{\Gamma(\tilde n_{\gamma}+|K/2|+1)}\right]^{1/2}
\label{24}
\end{equation}
The normalization constants for the $(n_{\gamma},K)=(0,0)$ and $(n_{\gamma},K)=(1,2)$ states are found to be
\begin{align}
N^2_{0,0}=2c, && N^2_{1,2}=6c^2
\label{25}
\end{align}
then $\frac{N^2_{1,2}}{N^2_{0,0}}=3c$, this result will be used to calculate the $B(E2)$ values in $\gamma \longrightarrow$ ground and $\gamma \longrightarrow \beta$ transitions ($\Delta K=2$).

In the prolate axial rotor case the energy spectrum is obtained as \cite{CLO15MR}
\begin{equation}
\epsilon _{n,L}=-\frac{1}{4}\left[\frac{\big(\alpha^2(\mu+n)^2-(A+2)^2\big)\big(\alpha^2(\mu+n)^2-A^2\big)}{\alpha^2(\mu+n)^2}\right]
\label{26}
\end{equation}
with
\begin{align}
\mu=&\frac{1}{2}+\sqrt{\frac{9}{4}+\Lambda+\frac{L(L+1)}{3}+\frac{A^2}{\alpha^2}}, \   A=e^{\alpha\beta_0}-1 \nonumber\\
\Lambda=&(6c)(n_{\gamma}+1)-\frac{K^2}{3}
\label{27}
\end{align}
To determine the radial wave function one needs the parametrization \cite{CLO15MR}
\begin{align}
\xi_{n,L}(\beta)&=\beta^{-2}\chi_{n,L}(\beta), && y=e^{-\alpha\beta}, &&  \nonumber \\ \chi_{n,L}(y)&=y^{\nu_{n,L}}(1-y)^\mu f_{n,L}(y), && \nu_{n,L}=\frac{\sqrt{1-\epsilon_{n,L}}}{\alpha},
\label{28}
\end{align}
leading to
\begin{align}
f''_{n,L}(y)=\frac{(1+2\mu+2\nu_{n,L})y-(1+2\nu_{n,L})}{y(1-y)}f'_{n,L}(y)
+\frac{\alpha^2(\nu_{n,L}+\mu)^2+\epsilon-e^{2\alpha\beta_0}}{\alpha^2y(1-y)}f_{n,L}(y)
\label{29}
\end{align}
The radial wave functions are found to be
 \begin{equation}
f_{n,L}(y)=N_{n,L \ 2 }F_1(-n,n+2\mu+2\nu_{n,L};1+2\nu_{n,L};y)  \label{30}
\end{equation}
while $_2F_1$ are hyper-geometrical functions, while $N_{n,L}$ are a normalization constant determined
from the normalization condition,
\begin{equation}
 \int_0^{\infty} \xi^2(\beta)\beta^4 d\beta=1  \label{31}
\end{equation}
with
\begin{align}
N_{n,L}=\left(\frac{\mu+n}{2\alpha n!\nu_{n,L}(\nu_{n,L}+\mu+n)} \right)^{-1/2}  \left( \frac{\big(\Gamma(2\nu_{n,L}+1)\Gamma(1+n)\big)^2}{\Gamma(2\nu_{n,L}+1+n)} \frac{\Gamma(2\mu+n)}{\Gamma(2\nu_{n,L}+2\mu+n)}\right)^{-1/2}  \label{32}
\end{align}
The $B(E2)$ transition rates for axially deformed nuclei around $\gamma=0$ read \cite{bijker03}
    \begin{align}
B(E2;nLn_{\gamma}K\longrightarrow n'L'n'_{\gamma}K')
=\frac{5}{16\pi}t^2\langle L,K,2,K'-K|L',K'\rangle^2 I^2_{n,L;n',L'}C^2_{n_{\gamma},K;n'_{\gamma},K'}
\label{34}
  \end{align}
with
\begin{align}
I_{n,L;n',L'}&=\int_0^{\infty} \beta \xi_{n',L'}(\beta)\xi_{n,L}(\beta)\beta^4d\beta \nonumber\\
&=\frac{1}{\alpha^2} \int_{-1}^{1} \left(\frac{\ln{\frac{1-z}{2}}}{1-z} \right) \chi_{n',L'}(z)\chi_{n,L}(z)   dz
\label{35}
  \end{align}
 $C_{_{n_{\gamma}K,n'_{\gamma}K'}}$ contains the integral over $\gamma$. For $\Delta K=0$ corresponding to transitions  ($g\rightarrow g, \gamma\rightarrow\gamma, \beta\rightarrow\beta$ and $\beta\rightarrow g $), the $\gamma$-integral part reduces to the orthonormality condition of the $\gamma$-wave functions : $C_{_{n_{\gamma},K;n'_{\gamma},K'}}=\delta_{_{n_{\gamma},n'_{\gamma}}}\delta_{_{K,K'}}$.
 While for $\Delta K=2$ corresponding to transitions ($\gamma\rightarrow g, \gamma\rightarrow\beta$), this
integral takes the form.
\begin{align}
C_{n_{\gamma},K;n'_{\gamma},K'}=\int^{\pi/3}_0 \sin\gamma\eta_{n_{\gamma}K}\eta_{n'_{\gamma}K'}|\sin3\gamma|d\gamma
\label{36}
  \end{align}
 Using the approximation $|\sin3\gamma| \approx 3|\gamma|$ and  Eq. \eqref{22} the last integral becomes
 \begin{align}
C_{n_{\gamma},K;n'_{\gamma},K'}=\frac{2 (3c)^{1+|K|/4+|K'|/4}}{\left(\Gamma(|K/2|+1)\Gamma(|K'/2|+1) \right)^{1/2}} \int^{\pi/3}_0 \gamma^{2+\frac{|K'|+|K|}{2}} e^{-3c\gamma^2} d\gamma
\label{37}
  \end{align}
where the Laguerre polynomials are unity since $\tilde n_{\gamma}=0$.
For the $(n'_{\gamma}=1,K'=2)\rightarrow(n_{\gamma},K=0)$ transition, Eq.\eqref{37}  is found to be
 \begin{equation}
C_{0,0;1,2}=\frac{1}{\sqrt{3c}}
\label{38}
  \end{equation}
\section{Numerical results \label{sec2}}

\begin{figure*}
\begin{center}
   \includegraphics[scale=0.86]{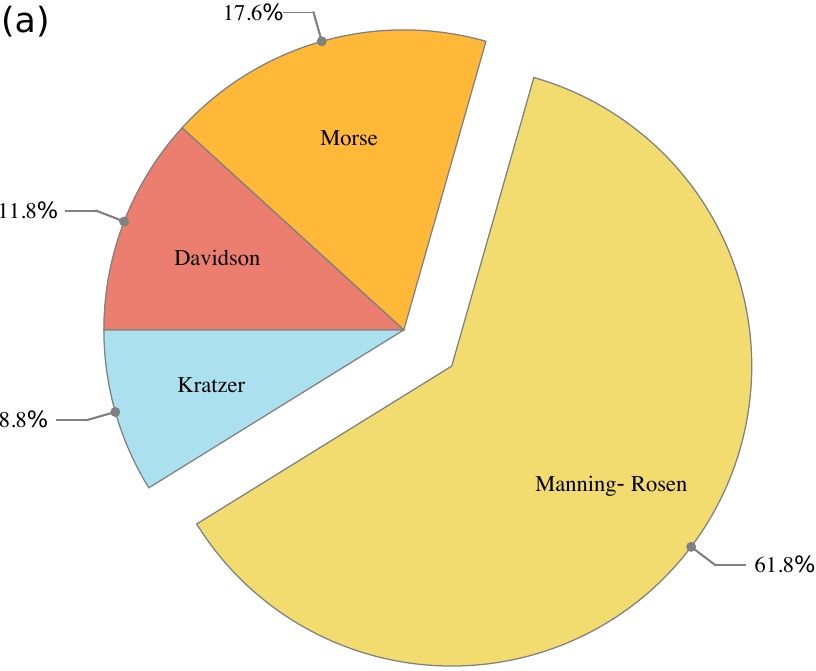}
   \includegraphics[scale=0.8]{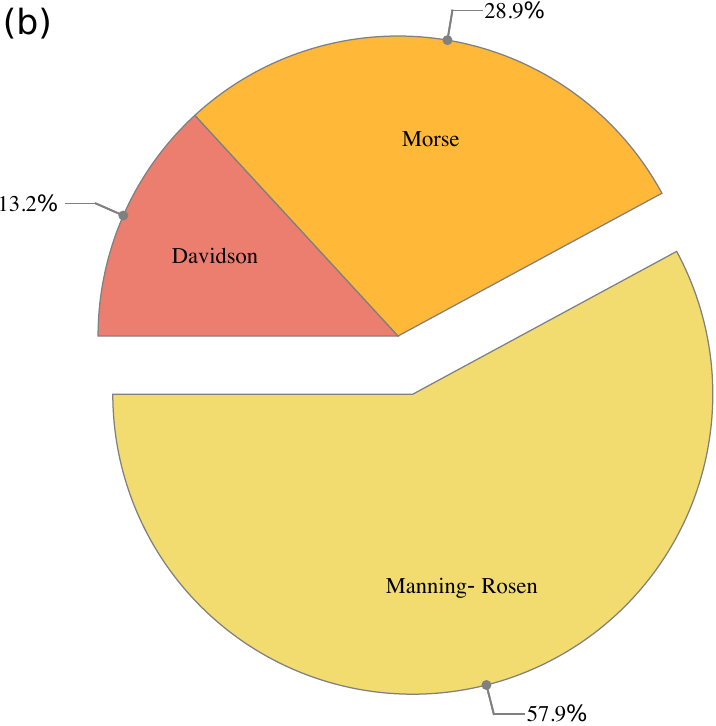}
   \caption{ (Color online) The percentage of the best value of the $rms$ \eqref{39} in Tables (\ref{tab3}-\ref{tab4}) obtained by Manning-Rosen $V_{MR}(\beta)$ \cite{CLO15MR},  Kratzer $V_K(\beta)$ \cite{bonat13}, Davidson $V_D(\beta)$ \cite{bonat11} and Morse $V_M(\beta)$ \cite{Inci11} potentials, in the $\gamma$-unstable case (left) and the rotational case (right).}\label{fig2}
\end{center}
\end{figure*}

In Table \eqref{tab1} and Table \eqref{tab2} are given the values of the free parameters of the present problem, namely : the Manning-Rosen potential parameters $\beta_0$ and $\alpha$ and the harmonic oscillator potential parameter $c$ for 34 $\gamma$-unstable and 38 axially symmetric prolate deformed nuclei respectively which are subject of the present study. These parameters have been already obtained in our previous work \cite{CLO15MR} by fitting the energy spectra to experimental data. We should notice here that such a fitting does not concern the transition probabilities.
In Table \eqref{tab3} and Table \eqref{tab4}, we present our  results for transition rates $B(E2)$ for $\gamma$-unstable and axially symmetric prolate deformed nuclei respectively. In the same tables are shown the experimental data for comparison. One can see, from those tables, that there is an overall agreement between our predictions and the experiment. Moreover, if we compare our results with those of Bonatsos et al.  obtained with Morse \cite{Inci11}, Davidson \cite{bonat11} or Kratzer \cite{bonat13} potentials, one can see from those tables  that our predictions are  generally better. This fact could be observed from the comparison of the root mean square (rms) deviation given by :

\begin{equation}
	rms = \sqrt{ \frac{1}{N} \sum_{j=1}^N  \left[\frac{B(E2,L_i \rightarrow L_f)_j^{(exp)}}{B(E2,2_g \rightarrow 0_g)^{(exp)}}-\frac{B(E2,L_i \rightarrow L_f)_j^{(th)}}{B(E2,2_g \rightarrow 0_g)^{(th)}}    \right]^2} .
	\label{39}
\end{equation}
where $N$ denotes  the total number of points involved in $rms$, while $B(E2,L_i \rightarrow L_f)_j^{(exp)}$ and $B(E2,L_i \rightarrow L_f)_j^{(th)}$ represent the theoretical and experimental transition rates from an initial to a final state, respectively. $B(E2,2_g \rightarrow 0_g)$ is the transition between the first-excited state and the ground state. It comes out from this comparison that over the 34 $\gamma$-unstable and 38 axially symmetric prolate deformed nuclei subject of this study, nearly $60\%$ are very well reproduced by the Manning-Rosen potential in both cases as can be seen from Figure \ref{fig2}.

\begin{figure*}
\begin{center}
 \includegraphics[width=0.49\textwidth]{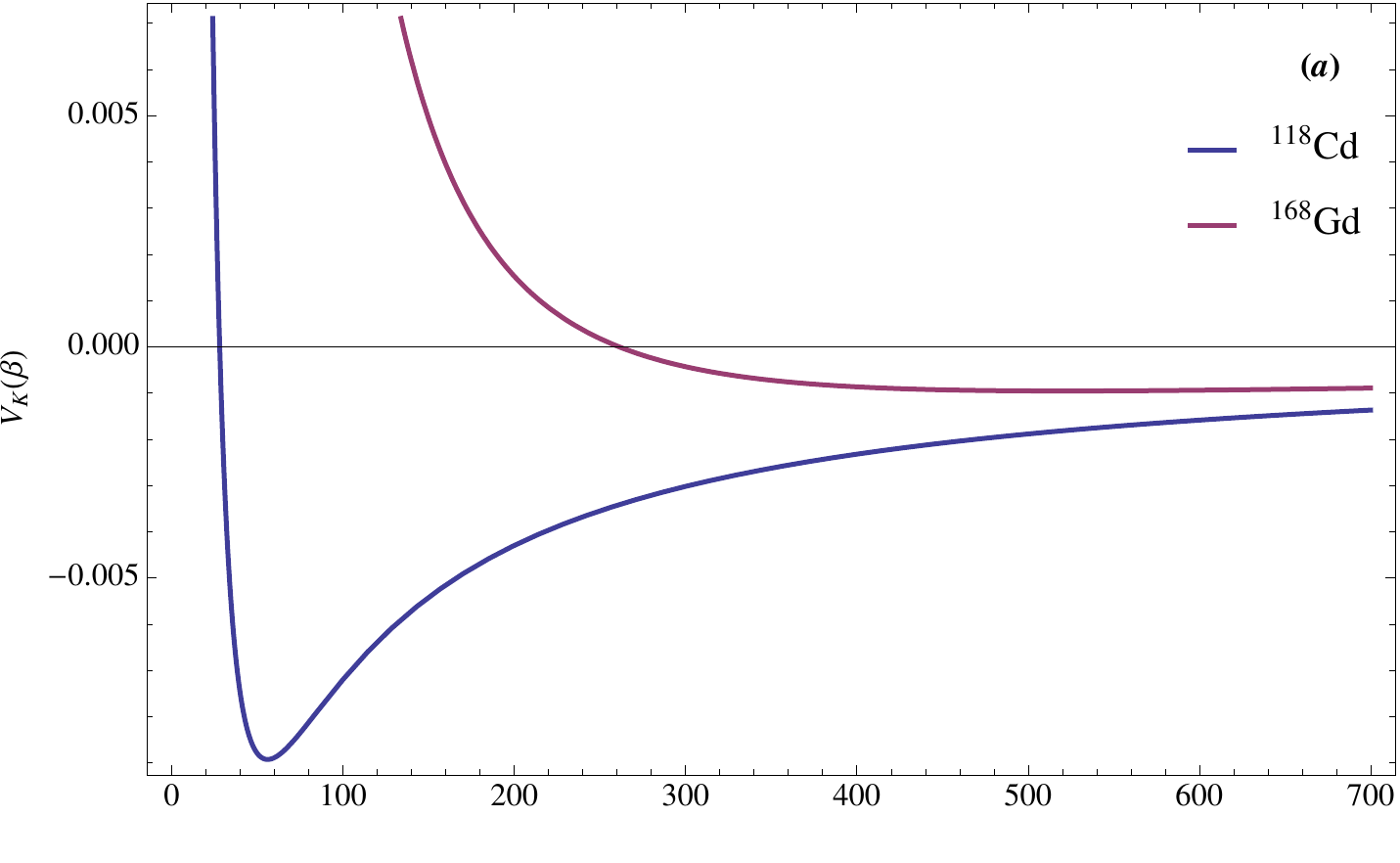}
\hspace{0.2cm}
	\includegraphics[width=0.47\textwidth]{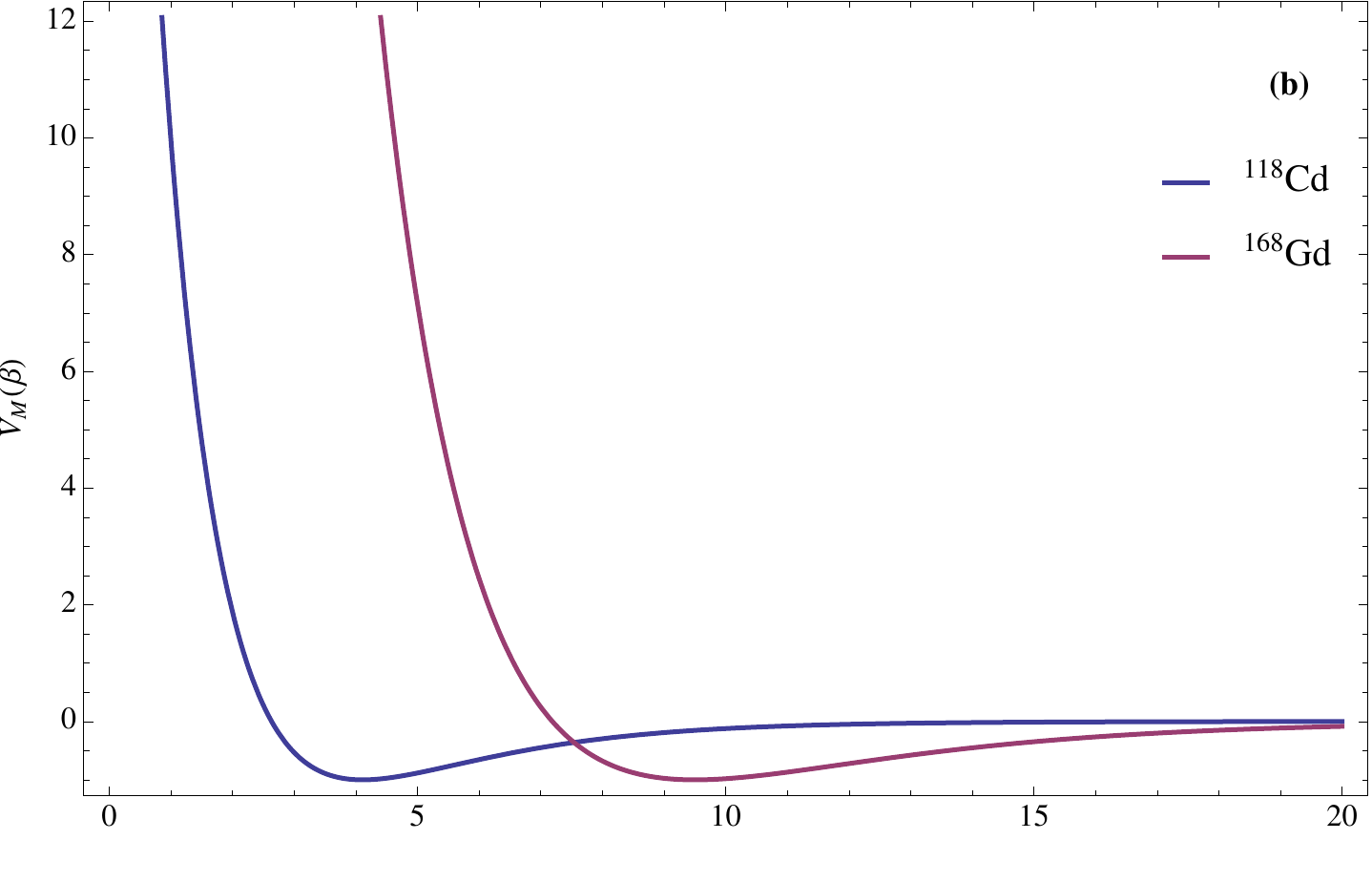}
\includegraphics[width=0.49\textwidth]{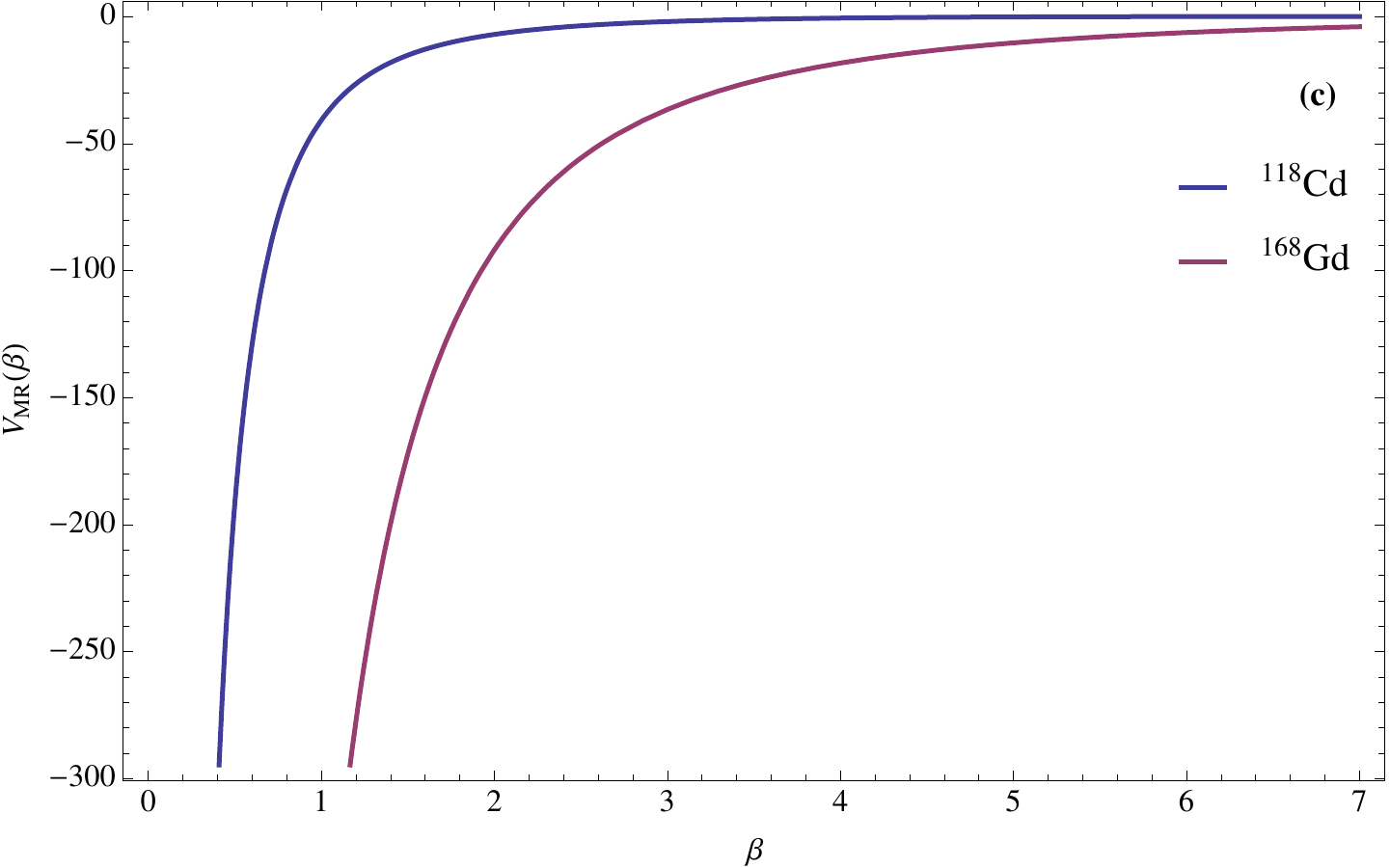}
\hspace{0.2cm}
	\includegraphics[width=0.47\textwidth]{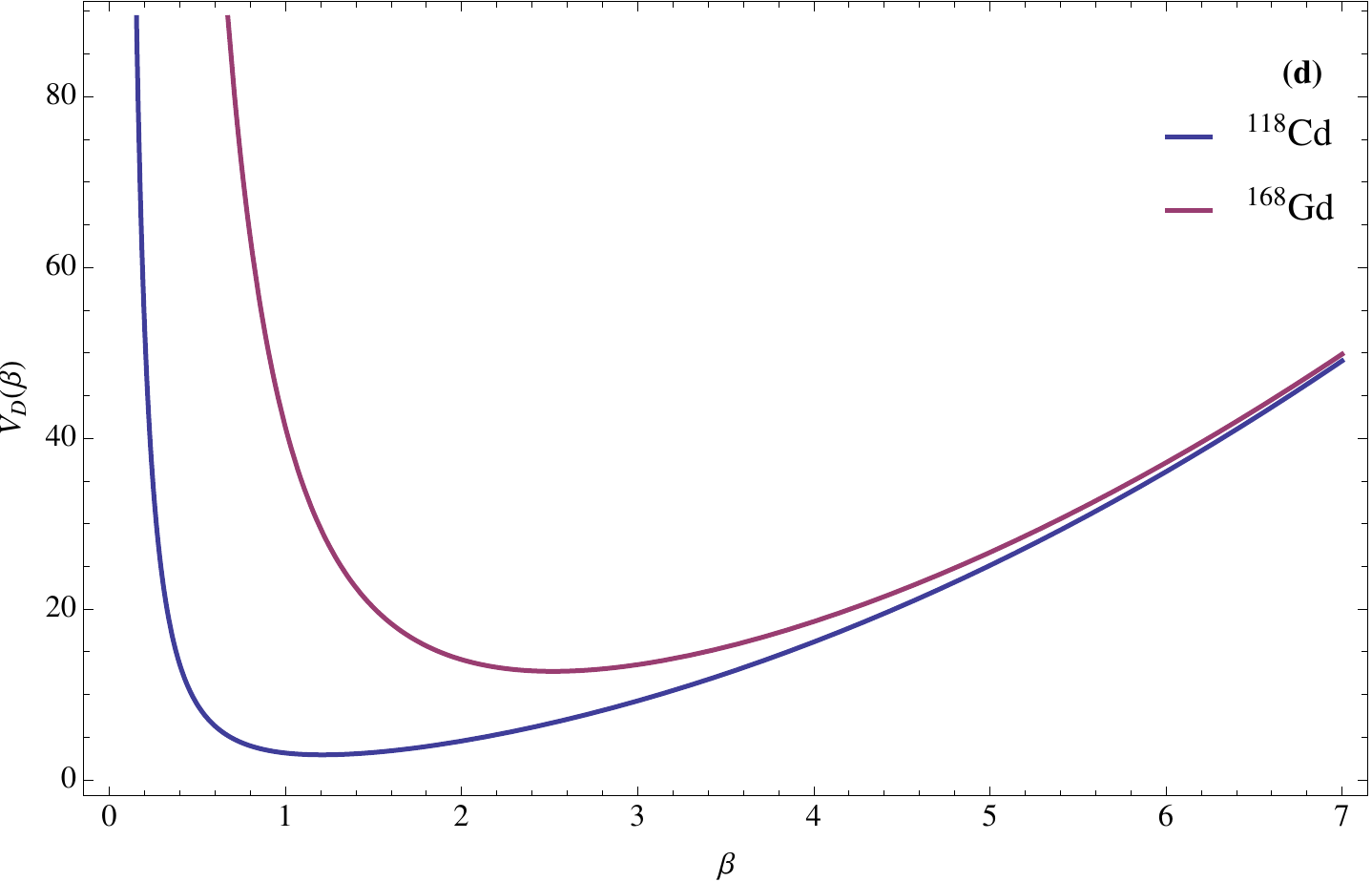}
   \caption{ (Color online) Evolution of   Kratzer $V_K(\beta)$ (a) \cite{bonat13}, Morse $V_M(\beta)$ (b)  \cite{Inci11}, Manning-Rosen $V_{MR}(\beta)$ (c)  \cite{CLO15MR} and  Davidson $V_D(\beta)$ (d) \cite{bonat11}   potentials, for $^{118}Cd$ and $^{162}Gd$ nuclei. The quantities shown are dimensionless. }\label{fig1}
\end{center}
\end{figure*}

The quality of our results in respect to those of Bonatsos et al. \cite{Inci11,bonat11,bonat13} is related to the shape of the used potential. Indeed, it has been shown \cite{Inci11} that if a considered potential increases rapidly for large values of $\beta$, this leads to large spacings in the $\beta$ band. From Figure \eqref{fig1}, one can observe that for large values of $\beta$, the Manning-Rosen potential becomes  flatter than Morse and Kratzer potentials, while the Davidson potential is growing as $\beta^2$. In addition, it should be stressed that the number of free parameters used in our calculations is the same as that used in Bonatsos et al. ones, namely : two parameters in the $\gamma$-unstable case and three parameters in the axially symmetric prolate deformed case. Also, the same experimental data have been used in both calculations.

In the case of $\gamma$-unstable nuclei, some inter-band transitions are systematically underestimated by all models presented in  Table \eqref{tab3}, while in the rotational nuclei some inter-band are systematically overestimated also by all models shown in Table \eqref{tab4}. But, in both cases our results are generally better in comparison with the other models.

\section{Conclusion \label{sec3}}
In the present work, we have extended our previous calculations \cite{CLO15MR} within the Bohr Hamiltonian with the Manning-Rosen potential as the $\beta$-part of the nuclear collective potential, while the $\gamma$-part was described by a harmonic oscillator potential, to include transition rates in the three bands, namely : the ground state, the $\beta$ and $\gamma$ bands of 34 $\gamma$-unstable and 38 axially symmetric prolate deformed nuclei. From the comparison of our  results with the experimental data and the theoretical predictions of other authors which have been obtained with Davidson, Kratzer and Morse potentials, it comes out that our results are better for nearly $60\%$ of the studied nuclei, while the calculation with  the other potentials succeeded to reproduce less than $30\%$ of such nuclei in the best case. So, it seems that the Manning-Rosen potential is more suitable for such calculations thanks to its shape which is flatter for large values of the $\beta$ collective coordinate in respect to the other potentials.
\bibliographystyle{IEEEtran}
\bibliography{B(E2)-Manuscript-MR}

\newpage
\begin{center}
\begin{table}
\caption{\label{tab1}The values of free parameters of Manning-Rosen potential used in the calculations for nuclei in the $\gamma$-unstable, taken from \cite{CLO15MR}}
\begin{tabular}{clc||clc||lcl||clclclclclclclcl}
\hline
\hline
nucleus   & $\beta_0$ & $\alpha$ & nucleus &  $\beta_0$ & $\alpha$& nucleus   & $\beta_0$ & $\alpha$& nucleus   & $\beta_0$ & $\alpha$\\
\hline

$^{  98}$Ru & 8.46&0.008&$   ^{106}$Cd&4.34&0.927&$   ^{122}$Xe&5.28&0.971& $   ^{152}$Gd&7.52&0.009\\
$   ^{100}$Ru& 11.60&0.009&$   ^{108}$Cd&8.74&0.016&$   ^{124}$Xe&5.18&0.959&$   ^{154}$Dy&10.44&0.01\\
$   ^{102}$Ru & 8.46&0.008&$   ^{110}$Cd&6.98&0.001&$   ^{128}$Xe&6.08&0.114& $   ^{156}$Er&9.95&0.001\\
$   ^{104}$Ru & 8.55&0.009&$   ^{112}$Cd&5.91&0.005&$   ^{132}$Xe&3.57&0.388&  $   ^{192}$Pt&4.54&0.926\\
$   ^{102}$Pd & 12.3&0.007&$   ^{114}$Cd&5.98&0.008&$   ^{130}$Ba&5.24&0.971&  $   ^{194}$Pt&6.80&0.102\\
$   ^{104}$Pd & 8.46&0.008&$   ^{116}$Cd&6.77&0.01&$   ^{132}$Ba&8.70&0.02&  $   ^{196}$Pt&4.55&0.924\\
$   ^{106}$Pd & 7.84&0.005&$   ^{118}$Cd&6.96&0.02&$   ^{134}$Ba&19.40&0.105&  $   ^{198}$Pt&4.33&0.160\\
$   ^{108}$Pd& 8.55&0.009&$   ^{118}$Xe&4.94&0.942&$   ^{142}$Ba&4.90&0.946\\
$   ^{110}$Pd& 0.00&0.001&$   ^{120}$Xe&5.55&0.970&$   ^{148}$Nd&8.73&0.017\\

\hline
\hline
\end{tabular}
\end{table}
\end{center}
\begin{center}
\begin{table}
\caption{\label{tab2}The values of free parameters of Manning-Rosen potential used in the calculations  for axially symmetric prolate deformed rare earth and actinide nuclei, taken from \cite{CLO15MR}}
\begin{tabular}{cccc||cccc||cccccccc}
\hline
\hline
nucleus   & $\beta_0$ & $\alpha$ &c& nucleus &  $\beta_0$ & $\alpha$&c& nucleus   & $\beta_0$ & $\alpha$&c&\\
\hline

$^{150}$Nd& 6.70&0.006&3.5&$^{166}$Er&$16.97$&$0.009$&$3.4$&$^{182}$W&16.23&0.002&4.4\\
$^{152}$Sm&$7.65$&$0.005$&$4.3$&$^{168}$Er&12.11&0.003&3.6&$^{184}$W&$9.95$&$0.001$&$3.1$\\
$^{154}$Sm&$14.03$&$0.003$&$7.2$&$^{170}$Er&0.00&0.001&3.5&$^{186}$W&0.00&0.001&1.9\\
 $^{154}$Gd&$11.37$&$0.002$&$2.9$&$^{166}$Yb&14.80&0.004&3.3&$^{186}$Os&10.39&0.002&2.4\\
 $^{156}$Gd& 16.24&0.002&5.0&$^{168}$Yb&21.05&0.002&3.8&$^{188}$Os&$8.61$&$0.006$&$1.5$\\
 $^{158}$Gd&$14.99$&$0.003$&$5.6$&$^{170}$Yb&15.95&0.003&5.4&$^{230}$Th&18.16&0.003&5.2\\
$^{156}$Dy & 9.08&0.004&2.5&$^{172}$Yb&14.89&0.001&7.2&$^{232}$Th&20.99&0.003&5.6\\
$^{158}$Dy & 15.28&0.004&3.4&$^{174}$Yb&21.76&0.001&7.7&$^{234}$U&22.55&0.003&7.6\\
$^{160}$Dy & 18.17&0.002&4.0&$^{176}$Yb&19.13&0.002&5.4&$^{236}$U&24.87&0.002&7.4\\
$^{162}$Dy &$18.17$ & $0.002$ &$4.0$&$^{174}$Hf&17.20&0.002&4.9&$^{238}$U&$26.64$&$0.001$&$8.6$\\
$^{164}$Dy &$22.86$&$0.003$&$3.4$&$^{176}$Hf&14.89&0.001&6.0&$^{238}$Pu &27.62&0.001&8.1\\
$^{162}$Er& 13.32&0.002&3.5&$^{178}$Hf&15.26&0.002&4.6&$^{250}$Cf&$25.68$&$0.005$&$8.4$\\
$^{164}$Er& 16.90&0.003&3.4&$^{180}$Hf&12.41&0.005&4.9&\\

\hline
\hline
\end{tabular}
\end{table}
\end{center}
\newpage
\footnotesize{
\begin{longtable}{lllllllllllllll}
\caption{
Comparison of experimental data (Exp) \cite{data}  for several B(E2) ratios of  $\gamma$-unstable nuclei to predictions by the Bohr Hamiltonian with the Manning-Rosen potential (MR), Morse potential (M) \cite{Inci11}, Davidson potential (D) \cite{bonat11} and Kratzer potential (K) \cite{bonat13}.  \label{tab3}}\\
\hline
nucl.&& $\frac{4_g\rightarrow 2_g}{2_g\rightarrow 0_g}$ & $\frac{6_g\rightarrow 4_g}{2_g\rightarrow 0_g}$ &$\frac{8_g\rightarrow 6_g}{2_g\rightarrow 0_g}$ & $\frac{10_g\rightarrow 8_g}{2_g\rightarrow 0_g}$ & $\frac{2_{\gamma}\rightarrow 2_g}{2_g\rightarrow 0_g}$&$\frac{2_{\gamma}\rightarrow 0_g}{2_g\rightarrow 0_g}$&$\frac{0_{\beta}\rightarrow 2_g}{2_g\rightarrow 0_g}$&$\frac{2_{\beta}\rightarrow 0_g}{2_g\rightarrow 0_g}$&rms\\
&&&&&&&$\times 10^3$&&$\times 10^3$  \\
\hline
\endfirsthead
\caption{(continued)}\\
\hline
nucl.&& $\frac{4_g\rightarrow 2_g}{2_g\rightarrow 0_g}$ & $\frac{6_g\rightarrow 4_g}{2_g\rightarrow 0_g}$ &$\frac{8_g\rightarrow 6_g}{2_g\rightarrow 0_g}$ & $\frac{10_g\rightarrow 8_g}{2_g\rightarrow 0_g}$ & $\frac{2_{\gamma}\rightarrow 2_g}{2_g\rightarrow 0_g}$&$\frac{2_{\gamma}\rightarrow 0_g}{2_g\rightarrow 0_g}$&$\frac{0_{\beta}\rightarrow 2_g}{2_g\rightarrow 0_g}$&$\frac{2_{\beta}\rightarrow 0_g}{2_g\rightarrow 0_g}$&rms\\
&&&&&&&$\times 10^3$&&$\times 10^3$  \\
\hline
\endhead
\hline
\endfoot
$   ^{  98}$Ru&Exp & $1.44(25)$ & $$&$ $ & $$&$1.62(61)$ & $36.0(152)$&$$&\\
 &MR& $1.58$ & $2.10$&$2.70 $ & $3.40$&$1.58$ & $0.0$&$0.24$&$11.50$&0.0498&\\
 &M& 1.71& 2.52 &3.61& 5.27 &1.71 &0.0 &0.82& 4.24&0.0942\\
 &D& 1.82 &2.62 &3.42 &4.22& 1.82 &0.0& 1.36& 3.60&0.1420\\
 &K& 1.77& 2.81 &4.63 &8.42 &1.77& 0.0& 1.27 &27.84&0.1198 \\\\

$   ^{100}$Ru& Exp& $1.45(13)$ & $$&$ $ & $$&$0.64(12)$ & $41.1(52)$&$0.98(15)$&&\\
 &MR& $1.52$ & $1.90$&$2.28 $ & $2.70$&$1.52$ & $0.0$&$0.18$&$14.53$&0.2980\\
&M& 1.65 &2.31& 3.07& 4.00 &1.656& 0.0 &0.72& 8.20&0.2669\\
 &D& 1.72 &2.40& 3.07 &3.73& 1.726& 0.0& 1.05& 10.89&0.2801\\
 &K& 1.70 &2.56 &3.93 &6.59 &1.70 &0.0& 1.11 &43.07&0.2740 \\\\

$   ^{102}$Ru& Exp& $1.50(24)$ & $$&$ $ & $$&$0.62(7)$ & $24.8(7)$&$0.80(14)$&&\\
 &MR& $1.58$ & $2.10$&$2.70 $ & $3.40$&$1.58$ & $0.0$&$0.24$&$11.50$&0.2773\\
&M& 1.73 &2.53 &3.54& 4.86& 1.73 &0.0& 0.88& 4.61&0.2823\\
 &D& 1.78& 2.54& 3.28 &4.01& 1.78& 0.0& 1.27& 8.70&0.3189\\
 &K&1.77 &2.82& 4.68& 8.67 &1.77& 0.0& 1.29& 31.06&0.3179 \\\\

$   ^{104}$Ru&Exp & $1.18(28)$ & $$&$ $ & $$&$0.63(15)$ & $35.0(84)$&$0.42(7)$&&\\
&MR & $1.58$ & $2.08$&$2.66 $ & $3.37$&$1.58$ & $0.0$&$0.23$&$11.72$&0.2619\\
 &M& 1.62& 2.10& 2.49& 2.80& 1.62& 0.0& 0.71 &16.34&0.2795\\
 &D& 1.63 &2.18& 2.71& 3.21 &1.63& 0.0& 0.79 &22.41&0.2883\\
 &K&1.60& 2.20& 2.95 &4.00& 1.60& 0.0 &0.68& 25.59&0.2713 \\\\

$   ^{102}$Pd&Exp & $1.56(19)$ & $$&$ $ & $$&$0.46(9)$ & $128.8(735)$&$$&&\\
 &MR& $1.50$ & $1.88$&$2.24 $ & $2.62$&$1.50$ & $0.0$&$0.17$&$14.71$&0.3470\\
 &M& 1.68 &2.35& 3.07& 3.86 &1.68 &0.0& 0.80& 8.20&0.4079\\
 &D& 1.76& 2.49& 3.19& 3.87& 1.76& 0.0& 1.22& 12.34&0.4375\\
 &K&1.63& 2.31 &3.25& 4.77& 1.63& 0.0& 0.87& 41.64&0.3900 \\\\

 $   ^{104}$Pd&Exp & $1.36(27)$ & $$&$ $ & $$&$0.61(8)$ & $33.3(74)$&$$&&\\
 &MR&$1.58$ & $2.10$&$2.70 $ & $3.40$&$1.58$ & $0.0$&$0.24$&$11.50$&0.3289\\
 &M& 1.67& 2.39 &3.30 &4.60 &1.67 &0.0& 0.73& 6.18&0.3655\\
 &D& 1.74& 2.45& 3.15& 3.85& 1.74& 0.0 &1.11& 8.13&0.3947\\
 &K&1.70 &2.52 &3.74& 5.83& 1.70& 0.0 &0.99 &24.16&0.3779 \\\\

 $   ^{106}$Pd&Exp & $1.63(28)$ & $$&$ $ & $$&$0.98(12)$ & $26.2(31)$&$0.67(18)$&&\\
 &MR& $1.60$ & $2.18$&$2.86 $ & $3.70$&$1.60$ & $0.0$&$0.26$&$10.15$&0.1861\\
 &M&1.70 &2.47& 3.51 &5.03& 1.70 &0.0& 0.79& 4.84&0.1830\\
 &D& 1.85& 2.67& 3.49 &4.28 &1.85& 0.0 &1.49& 5.98&0.3033\\
 &K&1.74 &2.66& 4.13 &6.83 &1.74 &0.0 &1.12& 22.91&0.2220 \\\\

 $   ^{108}$Pd&Exp & $1.47(20)$ & $2.16(28)$&$2.99(48) $ & $$&$1.43(14)$ & $16.6(18)$&$1.05(13)$&$1.09(29)$&\\
 &MR& $1.58$ & $2.08$&$2.66 $ & $3.37$&$1.58$ & $0.0$&$0.23$&$11.72$&0.1300\\
 &M&1.68& 2.32& 2.95& 3.53& 1.68& 0.0 &0.85& 9.45&0.0594\\
 &D&1.75 &2.45 &3.12& 3.75& 1.75& 0.0& 1.20& 15.82&0.0782\\
 &K&1.66 &2.38& 3.42& 5.11& 1.66& 0.0 &0.89& 30.31&0.0836 \\\\

 $   ^{110}$Pd&Exp & $1.71(34)$ & $$&$ $ & $$&$0.98(24)$ & $14.1(22)$&$0.64(10)$&$$&\\
 &MR& $1.28$ & $1.40$&$1.46 $ & $1.50$&$1.28$ & $0.0$&$4.04$&$860.20$&0.8599\\
 &D&1.76& 2.43& 3.01& 3.51& 1.76 &0.0 &1.31 &26.24&0.2567\\
 &K&1.60& 2.18& 2.94& 4.06& 1.60& 0.0 &0.75 &42.18& 0.1594\\\\

  $   ^{106}$Cd&Exp & $1.78(25)$ & $$&$ $ & $$&$0.43(12)$ & $93.0(127)$&$$&$$&\\
 & MR&$1.50$ & $1.88$&$2.31 $ & $2.90$&$1.50$ & $0.0$&$3.52$&$648.61$&0.3698\\
 &D&1.68 &2.32& 2.95 &3.58 &1.68& 0.0 &0.92& 10.44&0.4188\\
 &K&1.66& 2.37 &3.34 &4.76 &1.66& 0.0& 0.83 &16.97&0.4128 \\\\

  $   ^{108}$Cd&Exp & $1.54(24)$ & $$&$ $ & $$&$0.64(20)$ & $67.7(120)$&$$&$$&\\
 &MR& $1.56$ & $2.05$&$2.58 $ & $3.24$&$1.56$ & $0.0$&$0.22$&$12.37$&0.3069\\
 &M&1.69& 2.46& 3.48 &5.00& 1.69& 0.0& 0.77 &4.99&0.3535\\
 &D&1.85 &2.69& 3.52& 4.35 &1.85 &0.0 &1.49 &4.06&0.4161\\
 &K&1.67& 2.40 &3.43 &5.01 &1.67 &0.0& 0.87 &19.88&0.3460 \\\\

  $   ^{110}$Cd&Exp & $1.68(24)$ & $$&$ $ & $$&$1.09(19)$ & $48.9(78)$&$$&$9.85(595)$&\\
 & MR&$1.64$ & $2.32$&$3.15 $ & $4.26$&$1.64$ & $0.0$&$0.29$&$7.77$&0.1380\\
 &M&1.74& 2.62 &3.90& 6.05 &1.74& 0.0& 0.84& 2.76&0.1631\\
 &D&1.99& 2.97& 3.93 &4.87 &1.99 &0.0& 1.98 &1.61&0.2377\\
 &K&1.85 &3.14& 5.63 &11.54& 1.85& 0.0 &1.52 &20.99&0.1945 \\\\

$   ^{112}$Cd&Exp & $2.02(22)$ & $$&$ $ & $$&$0.50(10)$ & $19.9(35)$&$1.69(48)$&$11.26(210)$&\\
 &MR& $1.72$ & $2.53$&$3.62 $ & $5.10$&$1.72$ & $0.0$&$0.33$&$4.76$&0.3710\\
  &M&1.76 &2.70& 4.20& 4.97& 1.76& 0.0& 0.81& 1.56&0.3122\\
 &D&2.00& 2.99& 3.98& 4.96& 2.00 &0.0 &1.99 &0.48&0.3056\\
 &K&1.95& 3.53 &6.92& 15.92 &1.95 &0.0 &1.82 &12.87&0.2913\\\\

$   ^{114}$Cd&Exp & $1.99(25)$ & $3.83(72)$&$ 2.73(97)$ & $$&$0.71(24)$ & $15.4(29)$&$0.88(11)$&$10.61(193)$&\\
 &MR &$1.70$ & $2.50$&$3.55 $ & $4.98$&$1.70$ & $0.0$&$0.32$&$5.16$&0.2792\\
 &M&1.78& 2.79& 4.42 &3.44& 1.78 &0.0 &0.85 &0.99&0.3228\\
 &D&2.00& 2.99& 3.97& 4.94& 2.00& 0.0& 1.99& 0.74&0.3232\\
 &K&1.93 &3.46& 6.72& 15.44 &1.93 &0.0& 1.77& 15.44&0.6105 \\\\

$   ^{116}$Cd &Exp& $1.70(52)$ & $$&$ $ & $$&$0.63(46)$ & $32.8(86)$&$0.02$&$$&\\
 & MR& $1.64$ & $2.32$&$3.15 $ & $4.26$&$1.64$ & $0.0$&$0.29$&$7.75$&0.2609\\
 &M&1.66& 2.39& 3.43& 5.00& 1.66& 0.0& 0.63 &4.55&0.2986\\
 &D&1.74 &2.46& 3.17& 3.90& 1.74& 0.0& 1.11& 4.42&0.3883\\
 &K&1.69& 2.47& 3.52& 5.05& 1.69& 0.0& 0.90& 10.02&0.3437 \\\\

$   ^{118}$Cd&Exp & $1.85$ & $$&$ $ & $$&$$ & $$&$0.16$&$$&\\
 &MR &$1.63$ & $2.26$&$3.04 $ & $4.04$&$1.63$ & $0.0$&$0.28$&$8.60$&0.1253\\
 &M&1.66& 2.38 &3.42 &4.98 &1.66& 0.0 &0.62 &4.60&0.2488\\
 &D&1.71& 2.39& 3.06 &3.74& 1.71 &0.0 &1.00& 5.88&0.4258\\
 &K&1.70 &2.51 &3.65 &5.41 &1.70 &0.0 &0.95& 13.14&0.4021 \\\\

$   ^{118}$Xe&Exp & $1.11(7)$ & $0.88(27)$&$0.49(20) $ & $0.73$&$$ & $$&$$&$$&\\
 &MR& $1.46$ & $1.76$&$2.02 $ & $2.27$&$1.46$ & $0.0$&$3.29$&$628.38$&0.5913\\
 &M&1.65& 2.19& 2.65 &3.03 &1.65& 0.0& 0.78& 13.52&0.8639\\
 &D&1.67& 2.28& 2.85& 3.39& 1.67 &0.0& 0.95 &21.93&0.9648\\
 &K&1.61 &2.21& 3.00& 4.17& 1.61& 0.0& 0.74 &34.42&1.1216 \\\\

$   ^{120}$Xe&Exp & $1.16(14)$ & $1.17(24)$&$0.96(22) $ & $0.91(19)$&$$ & $$&$$&$$&\\
 &MR &$1.44$ & $1.72$&$1.91 $ & $2.08$&$1.44$ & $0.0$&$3.24$&$635.04$&0.4062\\
 &M&1.58 &2.05& 2.49& 2.90& 1.58& 0.0 &0.54 &15.06&0.6723\\
 &D&1.60& 2.11 &2.60& 3.08 &1.60 &0.0 &0.67 &21.51&0.7269\\
 &K&1.56 &2.06& 2.64 &3.43& 1.56& 0.0& 0.62& 42.25&0.7946 \\\\

$   ^{122}$Xe&Exp & $1.47(38)$ & $0.89(26)$&$0.44 $ & $$&$$ & $$&$$&$$&\\
 & MR&$1.44$ & $1.73$&$1.94 $ & $2.14$&$1.44$ & $0.0$&$3.32$&$646.12$&0.5727\\
 &D&1.58& 2.05& 2.48 &2.89& 1.58& 0.0& 0.63& 29.29&0.7826\\
 &K&1.54& 2.00 &2.52 &3.17 &1.54 &0.0& 0.54& 36.27&0.7858 \\\\

$   ^{124}$Xe&Exp & $1.34(24)$ & $1.59(17)$&$0.63(29) $ & $0.29(8)$&$0.70(19)$ & $15.9(46)$&$$&$$&\\
 &MR& $1.46$ & $1.73$&$1.96 $ & $2.18$&$1.46$ & $0.0$&$3.29$&$637.34$&0.4052\\
 &M&1.57& 2.05 &2.50& 2.92 &1.57& 0.0& 0.54 &14.87&0.5622\\
 &D&1.59 &2.09& 2.57& 3.04 &1.59& 0.0& 0.63& 20.14&0.5862\\
 &K&1.55 &2.03 &2.57& 3.25& 1.55& 0.0 &0.53& 28.40&0.6106 \\\\

$   ^{128}$Xe&Exp & $1.47(20)$ & $1.94(26)$&$2.39(40) $ & $2.74(114)$&$1.19(19)$ & $15.9(23)$&$$&$$&\\
 & MR&$1.60$ & $2.16$&$2.86 $ & $3.76$&$1.60$ & $0.0$&$0.25$&$10.10$&0.2031\\
 &M&1.60& 2.19& 2.90& 3.89& 1.60& 0.0 &0.55& 8.95&0.2249\\
 &D&1.63 &2.20 &2.75 &3.31& 1.63 &0.0 &0.73 &9.64&0.1428\\
 &K&1.83 &2.95 &4.73& 7.64& 1.83& 0.0& 0.75 &12.57&0.9281 \\\\

 $   ^{132}$Xe&Exp & $1.24(18)$ & $$&$ $ & $$&$1.77(29)$ & $3.4(7)$&$$&$$&\\
 & MR&$1.78$ & $2.80$&$0.36 $ & $0.34$&$1.78$ & $0.0$&$0.22$&$1.07$&0.1794\\
 &D&2.00 &3.00 &4.00 &5.00 &2.00 &0.0 &2.00 &0.00&0.2638\\
 &K&2.78 &7.13& 17.89 &43.35& 2.78& 0.0& 2.49 &0.07& 0.6129\\\\

 $   ^{130}$Ba&Exp & $1.36(6)$ & $1.62(15)$&$1.55(56) $ & $0.93(15)$&$$ & $$&$$&$$&\\
 & MR&$1.46$ & $1.73$&$1.94 $ & $2.14$&$1.46$ & $0.0$&$3.33$&$647.59$&0.3191\\
 &M&1.66& 2.28& 2.90 &3.50 &1.66 &0.0& 0.78& 10.07&0.7469\\
 &D&1.56 &2.01& 2.41& 2.77 &1.56& 0.0& 0.61 &34.54&0.5183\\
 &K&1.54& 2.01& 2.54& 3.22& 1.54& 0.0& 0.56& 39.43&0.6318 \\\\

 $   ^{132}$Ba&Exp & $$ & $$&$ $ & $$&$3.35(64)$ & $90.7(177)$&$$&$$&\\
 &MR &$1.56$ & $2.05$&$2.56 $ & $3.20$&$1.56$ & $0.0$&$0.22$&$12.50$&0.8993\\
  &M&1.48& 1.84& 2.19 &2.66 &1.48& 0.0 &0.00& 0.00&0.9393\\
 &D&1.68 &2.30& 2.90 &3.50& 1.68& 0.0 &0.92 &15.21&0.8394\\
 &K&1.61& 2.20& 2.94& 3.95& 1.61 &0.0& 0.66& 20.59& 0.8744\\\\

  $   ^{134}$Ba&Exp & $1.55(21)$ & $$&$ $ & $$&$2.17(69)$ & $12.5(41)$&$$&$$&\\
 &MR& $1.43$ & $1.68$&$1.86 $ & $1.98$&$1.43$ & $0.0$&$0.05$&$8.86$&0.2523\\
  &M&1.70& 2.53& 3.81& 1.96 &1.70 &0.0& 0.68& 2.48&0.1665\\
 &D&1.75& 2.48& 3.21 &3.94& 1.75 &0.0& 1.14 &4.08&0.1569\\
 &K&2.13& 4.10& 7.88& 15.19& 2.13 &0.0& 1.26 &6.22&0.1933 \\\\

  $   ^{142}$Ba&Exp & $1.40(17)$ & $0.56(14)$&$ $ & $$&$$ & $$&$$&$$&\\
 &MR& $1.46$ & $1.76$&$2.02 $ & $2.29$&$1.46$ & $0.0$&$3.33$&$635.55$&0.6000\\
 &M&1.50 &1.89& 2.30& 2.88& 1.50& 0.0& 0.11& 2.37&0.6661\\
 &D&1.55 &2.00& 2.41& 2.82& 1.55& 0.0& 0.49& 18.60&0.7231\\
 &K&1.54 &1.99& 2.46& 3.04 &1.54 &0.0 &0.45& 21.34&0.7176 \\\\

 $   ^{148}$Nd&Exp & $1.61(13)$ & $1.76(19)$&$ $ & $$&$0.25(4)$ & $9.3(17)$&$0.54(6)$&$32.82(816)$&\\
 &MR &$1.56$ & $2.05$&$2.58 $ & $3.22$&$1.56$ & $0.0$&$0.22$&$12.41$&0.2296\\
 &M&1.63& 2.14& 2.56 &2.89& 1.63& 0.0& 0.73 &15.09&0.2398\\
 &D&1.63& 2.17& 2.68 &3.15 &1.63 &0.0& 0.81& 26.86&0.2432\\
 &K&1.57& 2.08& 2.67& 3.47 &1.57& 0.0& 0.59& 30.88&0.2259 \\\\

 $   ^{152}$Gd &Exp& $1.84(29)$ & $2.74(81)$&$ $ & $$&$0.23(4)$ & $4.2(8)$&$2.47(78)$&$$&\\
 & MR&$1.62$ & $2.20$&$2.91 $ & $3.81$&$1.62$ & $0.0$&$0.26$&$9.67$&0.5363\\
  &M&1.71& 2.47& 3.39& 4.54& 1.71& 0.0 &0.85& 5.63&0.4438\\
 &D&1.98 &2.92 &3.81& 4.65& 1.98& 0.0& 1.95& 4.51&0.3674\\
 &K&1.80 &2.96 &5.14 &10.30& 1.80 &0.0 &1.41 &32.70&0.3815 \\\\

  $   ^{154}$Dy&Exp & $1.62(35)$ & $2.05(42)$&$2.27(62) $ & $1.86(69)$&$$ & $$&$$&$$&\\
 &MR& $1.53$ & $1.95$&$2.38 $ & $2.87$&$1.53$ & $0.0$&$0.19$&$13.92$&0.2545\\
  &M&1.67& 2.37 &3.20 &4.25 &1.67& 0.0& 0.76& 7.08&0.6439\\
 &D&1.91 &2.79& 3.64& 4.46& 1.91& 0.0 &1.70& 5.41&0.7586\\
 &K&1.78 &2.89& 5.06 &10.73& 1.78& 0.0 &1.46& 58.09&2.3322 \\\\

  $   ^{156}$Er &Exp& $1.78(16)$ & $1.89(36)$&$0.76(20) $ & $0.88(22)$&$$ & $$&$$&$$&\\
 &MR &$1.54$ & $2.00$&$2.50 $ & $3.06$&$1.54$ & $0.0$&$0.21$&$13.12$&0.6997\\
  &M&1.62& 2.23 &2.91 &3.74 &1.62& 0.0& 0.64 &9.60&0.8986\\
 &D&1.70 &2.35 &3.00& 3.64& 1.70& 0.0& 0.98 &11.50&0.8954\\
 &K&1.64& 2.33 &3.27& 4.73& 1.64& 0.0 &0.83 &28.76&1.1540 \\\\

   $   ^{192}$Pt&Exp & $1.56(12)$ & $1.23(55)$&$ $ & $$&$1.91(16)$ & $9.5(9)$&$$&$$&\\
 &MR& $1.48$ & $1.83$&$2.18 $ & $2.62$&$1.48$ & $0.0$&$3.40$&$632.98$&0.1848\\
  &M&1.58 &2.08& 2.55& 3.00& 1.58 &0.0& 0.56 &14.25&0.2269\\
 &D&1.59& 2.09 &2.57 &3.05 &1.59 &0.0 &0.61& 16.98&0.2284\\
 &K&1.57 &2.09& 2.68& 3.44& 1.57 &0.0 &0.54 &17.79&0.2301 \\\\

   $   ^{194}$Pt &Exp& $1.73(13)$ & $1.36(45)$&$1.02(30) $ & $0.69$&$1.81(25)$ & $5.9(9)$&$0.01$&$$&\\
 & MR&$1.58$ & $2.06$&$2.64 $ & $3.35$&$1.58$ & $0.0$&$0.22$&$11.82$&0.4564\\
  &M&1.58& 2.06 &2.49& 2.88 &1.58 &0.0 &0.58 &15.34&0.3829\\
 &D&1.59 &2.09& 2.57 &3.04& 1.59& 0.0& 0.63 &19.78&0.4073\\
 &K&1.56& 2.07 &2.63& 3.34& 1.56& 0.0& 0.52& 19.45& 0.4503\\\\

    $   ^{196}$Pt&Exp & $1.48(3)$ & $1.80(23)$&$1.92(23) $ & $$&$$ & $0.4$&$0.07(4)$&$0.06(6)$&\\
 &MR &$1.48$ & $1.83$&$2.18$ & $2.62$&$1.48$ & $0.0$&$3.38$&$629.69$&0.5626\\
 &M&1.63& 2.14& 2.57 &2.93 &1.63& 0.0 &0.72 &14.74&0.1644\\
 &D&1.64 &2.21 &2.75& 3.28& 1.64 &0.0 &0.82& 20.83&0.1995\\
 &K&1.61 &2.21 &2.97& 4.04& 1.61& 0.0& 0.69 &23.11& 0.2147\\\\

     $   ^{198}$Pt &Exp& $1.19(13)$ & $1.78$&$ $ & $$&$1.16(23)$ & $1.2(4)$&$0.81(22)$&$1.56(126)$&\\
 & MR&$1.74$ & $2.64$&$4.00$ & $6.20$&$1.74$ & $0.0$&$0.33$&$3.27$&0.2113\\
 &M&1.72 &2.48& 3.35& 4.36 &1.72& 0.0& 0.89& 5.78&0.1737\\
 &D&1.82 &2.60 &3.36 &4.08 &1.82& 0.0 &1.41& 10.09&0.2272\\
 &K&1.76& 2.73 &4.24& 6.76& 1.76& 0.0 &1.16& 11.09&0.2176 \\

\hline
\end{longtable}
}

\newpage
\scriptsize{
\begin{longtable}{lllllllllllllllll}
\caption{ Comparison of experimental data \cite{data} (upper line) for several B(E2) ratios of axially symmetric prolate deformed nuclei to predictions (lower line) by the Bohr Hamiltonian with the Manning-Rosen potential (MR), Morse potential (M) \cite{Inci11}, Davidson potential (D) \cite{bonat11} and Kratzer potential (K) \cite{bonat13}.
\label{tab4}}\\
\hline
nucl. && $\frac{4_g\rightarrow 2_g}{2_g\rightarrow 0_g}$ & $\frac{6_g\rightarrow 4_g}{2_g\rightarrow 0_g}$ &$\frac{8_g\rightarrow 6_g}{2_g\rightarrow 0_g}$ & $\frac{10_g\rightarrow 8_g}{2_g\rightarrow 0_g}$ & $\frac{2_{\beta}\rightarrow 0_g}{2_g\rightarrow 0_g}$&$\frac{2_{\beta}\rightarrow 2_g}{2_g\rightarrow 0_g}$&$\frac{2_{\beta}\rightarrow 4_g}{2_g\rightarrow 0_g}$&$\frac{2_{\gamma}\rightarrow0_g}{2_g\rightarrow 0_g}$&$\frac{2_{\gamma}\rightarrow 2_g}{2_1\rightarrow 0_g}$&$\frac{2_{\gamma}\rightarrow 4_g}{2_g\rightarrow 0_g}$&rms\\
&&&&&&$\times 10^3$&$\times 10^3$&$\times 10^3$&$\times 10^3$&$\times 10^3$&$\times 10^3$&\\
\hline
\endfirsthead
\caption{(continued)}\\
\hline
nucl. && $\frac{4_g\rightarrow 2_g}{2_g\rightarrow 0_g}$ & $\frac{6_g\rightarrow 4_g}{2_g\rightarrow 0_g}$ &$\frac{8_g\rightarrow 6_g}{2_g\rightarrow 0_g}$ & $\frac{10_g\rightarrow 8_g}{2_g\rightarrow 0_g}$ & $\frac{2_{\beta}\rightarrow 0_g}{2_g\rightarrow 0_g}$&$\frac{2_{\beta}\rightarrow 2_g}{2_g\rightarrow 0_g}$&$\frac{2_{\beta}\rightarrow 4_g}{2_g\rightarrow 0_g}$&$\frac{2_{\gamma}\rightarrow0_g}{2_g\rightarrow 0_g}$&$\frac{2_{\gamma}\rightarrow 2_g}{2_1\rightarrow 0_g}$&$\frac{2_{\gamma}\rightarrow 4_g}{2_g\rightarrow 0_g}$&rms\\
&&&&&&$\times 10^3$&$\times 10^3$&$\times 10^3$&$\times 10^3$&$\times 10^3$&$\times 10^3$&\\
\hline
\endhead
\hline
\endfoot
$^{150}$Nd&Exp & $1.52(4)$ & $1.84(14)$&$ 2.05(13)$ & $$&$4.4(8)$ & $61.7(98)$&$174(55)$&$26.1(22)$&$49.6(26)$&$14.8(98)$\\
 &MR& $1.53$ & $1.91$&$2.36 $ & $2.97$&$25.36$ & $69.51$&$349.03$&$6.25$&$136.16$&$8.07$&0.2176\\
  &K&1.55& 1.98& 2.55& 3.40& 36.4& 93.5& 443& 95.9& 150.0& 9.0&0.0664\\\\

$^{152}$Sm &Exp&$1.45(5)$&$1.70(7)$&$1.98(14)$&$2.22(25)$&$6.4(7)$&$38.2(43)$&$132(15)$&$25.1(17)$&$64.6(48)$&$5.4(5)$\\
 & MR&$1.51$ & $1.84$&$2.22$&$2.70$&$25.17$ & $62.89$&$284.15$&$69.86$&$107.33$&$6.22$&0.0566\\
  &K&1.55 &1.98& 2.58 &3.53& 49.1 &113.4& 475& 84.0& 130.9& 7.8&0.1501\\\\

$^{154}$Sm&Exp &$1.40(5)$&$1.67(7)$&$1.83(11)$&$1.81(11)$&$5.4(13)$&$$&$25(6)$&$18.4(34)$&$$&$3.9(7)$\\
 & MR&$1.46$ & $1.68$&$1.86$&$2.05$&$20.83$ & $40.11$&$127.16$&$43.52$&$63.90$&$3.39$&0.0336\\
 &M& 1.46& 1.67 &1.85 &2.03& 21.1& 39.8 &122& 47.3& 69.2& 3.6&0.0311\\
  &D&1.47 &1.69& 1.87& 2.06 &26.7 &50.0& 150 &47.5& 69.6& 3.7&0.0364\\
  &K&1.46 &1.67 &1.86& 2.05 &24.7& 45.7 &136 &47.8 &69.9& 3.7&0.0341\\\\

 $^{154}$Gd&Exp &$1.56(7)$&$1.82(11)$&$1.99(12)$&$2.29(27)$&$5.5(5)$&$42.7(41)$&$125(11)$&$36.3(34)$&$78.3(69)$&$11.0(10)$\\
 & MR&$1.48$ & $1.74$&$1.99$&$2.28$&$23.43$ & $50.01$&$184.53$&$116.96$&$172.02$&$9.12$&0.0183\\
  &K&1.55& 1.98 &2.57& 3.54& 55.4& 122.5 &486& 114.7 &175.6& 10.1&0.1439\\\\

$^{156}$Gd &Exp&$1.41(5)$&$1.58(6)$&$1.71(10)$&$1.68(9)$&$3.4(3)$&$18(2)$&$22(2)$&$25.0(15)$&$38.7(24)$&$4.1(3)$\\
 & MR&$1.46$ & $1.66$&$1.82$&$1.99$&$19.71$ & $36.77$&$110.87$&$67.06$&$97.41$&$5.05$&0.0352\\
 &M& 1.47& 1.70 &1.90& 2.12& 22.5 &44.3& 145& 63.0& 92.4& 4.9&0.0508\\
  &D&1.48 &1.73 &1.95 &2.18 &29.7& 59.1& 191& 62.5& 92.4 &4.9&0.0599\\
  &K&1.47& 1.69 &1.90& 2.13& 30.8& 56.5& 166 &70.7 &103.3& 5.4&0.0523\\\\

$^{158}$Gd&Exp &$1.46(5)$&$$&$1.67(16)$&$1.72(16)$&$1.6(2)$&$0.4(1)$&$7.0(8)$&$17.2(20)$&$30.3(45)$&$1.4(2)$\\
  &MR&$1.46$ & $1.67$&$1.84$&$2.03$&$20.39$ & $38.75$&$120.37$&$58.94$&$85.98$&$4.50$&0.0420\\
  &M& 1.45 &1.64 &1.79 &1.95 &14.6& 26.7& 79& 63.3& 91.7 &4.7&0.0311\\
  &D&1.46 &1.66 &1.82& 1.98& 25.7& 45.9& 127& 64.0& 93.0& 4.8&0.0372\\
  &K&1.45 &1.66 &1.82 &1.98 &24.1& 42.8 &119& 63.9 &92.6& 4.8&0.0368\\\\

$^{156}$Dy &Exp&$1.75(14)$&$1.34(12)$&$1.94(13)$&$2.45(21)$&$$&$$&$$&$48.2(35)$&$63.0(78)$&$84.4(141)$\\
  &MR &$1.50$ & $1.81$&$2.14$&$2.56$&$24.86$ & $59.14$&$251.74$&$135.31$&$201.86$&$10.99$&0.0865\\
  &K&1.56 &2.03 &2.70& 3.83& 53.2& 124.3& 531& 151.8& 232.6& 13.4&0.2487\\\\

$^{158}$Dy&Exp &$1.45(10)$&$1.86(12)$&$1.86(38)$&$1.75(28)$&$12(3)$&$19(4)$&$66(16)$&$32.2(78)$&$103.8(258)$&$11.5(48)$\\
 & MR&$1.46$ & $1.67$&$1.85$&$2.03$&$20.40$ & $38.78$&$120.48$&$99.94$&$145.02$&$7.49$&0.0351\\
 &M&1.48 &1.73& 1.98 &2.26& 23.5& 49.0 &175& 95.1& 140.1& 7.5&0.0554\\
  &D&1.50 &1.78 &2.04& 2.31 &30.5& 65.4 &232& 88.5& 131.7& 7.1&0.0620\\
  &K&1.48 &1.73& 1.98& 2.28 &32.5 &63.0& 202 &97.9 &143.6 &7.6&0.0580\\\\

 $^{160}$Dy&Exp &$1.46(7)$&$1.23(7)$&$1.70(16)$&$1.69(9)$&$3.4(4)$&$$&$8.5(10)$&$23.2(21)$&$43.8(42)$&$3.1(3)$\\
 & MR&$1.45$ & $1.65$&$1.79$&$1.94$&$18.59$ & $33.73$&$97.21$&$84.80$&$122.55$&$6.28$&0.0561\\
  &M&1.46 &1.67 &1.84& 2.02& 20.9& 39.3 &120& 82.7& 120.1& 6.2&0.0640\\
  &D&1.46& 1.68& 1.85 &2.03& 22.9& 43.5& 133 &78.6 &114.5& 6.0&0.0659\\
  &K&1.46 &1.66 &1.83& 2.00 &23.5 &42.5& 122& 87.4 &126.6& 6.5&0.0619\\\\

$^{162}$Dy&Exp &$1.45(7)$&$1.51(10)$&$1.74(10)$&$1.76(13)$&$$&$$&$$&$0.12(1)$&$0.20$&$0.02$\\
 & MR&$1.45$ & $1.65$&$1.79$&$1.94$&$18.59$ & $33.73$&$97.21$&$84.80$&$122.55$&$6.28$&0.0394\\
 &M&1.45 &1.62& 1.76& 1.88& 11.4& 20.1& 56 &94.3& 135.6 &6.9&0.0331\\
  &D&1.45 &1.65& 1.80& 1.95& 23.9& 42.4& 116& 89.8& 129.8 &6.7&0.0412\\
  &K&1.45& 1.65 &1.80& 1.95 &23.7& 41.4 &112& 92.3 &133.2& 6.8&0.0416\\\\

$^{164}$Dy&Exp &$1.30(7)$&$1.56(7)$&$1.48(9)$&$1.69(9)$&$$&$$&$$&$19.1(22)$&$38.3(39)$&$4.6(5)$\\
 &MR& $1.44$ & $1.61$&$1.74$&$1.85$&$15.91$ & $27.35$&$71.78$&$99.64$&$143.21$&$7.26$&0.0501\\
 &M&1.44 &1.62 &1.74 &1.85 &16.1& 27.6& 72 &99.7& 143.2 &7.3&0.0503\\
  &D&1.44 &1.62 &1.75 &1.86 &16.9& 29.1 &77& 99.7 &143.4& 7.3&0.0519\\
  &K&1.45 &1.64 &1.79& 1.93& 23.6& 40.6& 107& 100.4& 144.7 &7.4&0.0619\\\\

$^{162}$Er&Exp &$$&$$&$$&$$&$8(7)$&$$&$170(90)$&$32.5(28)$&$77.0(56)$&$9.4(69)$\\
 & MR&$1.47$ & $1.70$&$1.90$&$2.13$&$21.93$ & $43.82$&$146.95$&$96.69$&$141.23$&$7.40$&0.0190\\
 &M&1.47 &1.72& 1.95& 2.21& 22.0 &45.1 &158& 96.6 &141.7 &7.5&0.0186\\
  &D&1.49 &1.75 &1.99 &2.24& 27.8& 58.3& 202 &91.1 &134.8& 7.2&0.0180\\
  &K&1.48 &1.73 &1.97 &2.25& 28.9& 57.1& 189& 100.4 &147.1 &7.8&0.0202\\\\

 $^{164}$Er&Exp &$1.18(13)$&$$&$1.57(9)$&$1.64(11)$&$$&$$&$$&$23.9(35)$&$52.3(72)$&$7.8(12)$\\
 & MR&$1.46$ & $1.66$&$1.81$&$1.98$&$19.36$ & $35.81$&$106.42$&$99.97$&$144.61$&$7.42$&0.0850\\
  &M&1.47 &1.69 &1.87& 2.05& 23.4 &44.8 &139 &103.3& 150.3 &7.8&0.0988\\
  &D&1.47& 1.70 &1.89& 2.09& 28.3& 53.5 &162& 103.8 &151.2 &7.9&0.1051\\
  &K&1.46 &1.67& 1.86 &2.05& 27.0& 49.0& 141& 110.5 &160.2& 8.2&0.0976\\\\

$^{166}$Er &Exp&$1.45(12)$&$1.62(22)$&$1.71(25)$&$1.73(23)$&$$&$$&$$&$25.7(31)$&$45.3(54)$&$3.1(4)$\\
 & MR&$1.45$ & $1.65$&$1.80$&$1.95$&$18.77$ & $34.22$&$99.34$&$99.95$&$144.37$&$7.39$&0.0381\\
 &M&1.45 &1.64 &1.78& 1.92& 17.3& 30.9& 88& 104.5& 150.7& 7.7&0.0342\\
  &D&1.46 &1.66 &1.81 &1.96& 20.7& 38.2& 111 &100.0 &144.8& 7.4&0.0400\\\\

$^{168}$Er&Exp &$1.54(7)$&$2.13(16)$&$1.69(11)$&$1.46(11)$&$$&$$&$$&$23.2(15)$&$41.1(31)$&$3.0(3)$\\
 & MR&$1.45$ & $1.65$&$1.80$&$1.94$&$18.63$ & $33.83$&$97.64$&$94.35$&$136.28$&$6.98$&0.1007\\
  &M&1.44 &1.61 &1.73& 1.84 &1.8& 3.1& 8& 101.1& 145.1& 7.3&0.0953\\
  &D&1.45 &1.65 &1.79 &1.93& 27.7& 47.2& 120& 100.6& 145.1& 7.4&0.0998\\
  &K&1.45 &1.64& 1.78 &1.92& 27.6& 46.2& 116& 100.6& 144.9& 7.4&0.0996\\\\

 $^{170}$Er &Exp&$$&$$&$1.78(15)$&$1.54(11)$&$1.4(1)$&$0.2(2)$&$6.8(12)$&$17.7(9)$&$$&$1.4(4)$\\
 & MR&$1.41$ & $1.53$&$1.57$&$1.58$&$645.85$ & $913.53$&$1608.76$&$91.50$&$130.70$&$6.53$&0.2809\\
 &M&1.46 &1.68 &1.84 &1.98 &28.1 &50.6& 139 &76.1 &110.6 &5.7&0.0670\\
  &D&1.47 &1.69& 1.86& 2.03& 39.2 &67.9 &177 &78.6 &114.2 &5.9&0.0761\\
  &K&1.46 &1.66 &1.83 &2.01 &42.8 &70.7 &173 &84.6 &122.2& 6.3&0.0730\\\\

 $^{166}$Yb&Exp &$1.43(9)$&$1.53(10)$&$1.70(18)$&$1.61(80)$&$$&$$&$$&$$&$$&$$\\
 &MR& $1.46$ & $1.68$&$1.86$&$2.05$&$20.75$ & $39.85$&$125.85$&$103.00$&$149.60$&$7.74$&0.1210\\
  &M&1.48& 1.73& 1.97 &2.23 &24.3 &50.3 &176 &101.4& 149.0 &7.9&0.1746\\
  &D&1.50& 1.78 &2.05 &2.33 &33.7 &71.0 &245 &97.2 &144.5 &7.8&0.2083\\
  &K&1.48& 1.73 &1.97 &2.27 &35.4 &66.7 &206 &115.2 &168.2 &8.8&0.1834\\\\

 $^{168}$Yb&Exp &$$&$$&$$&$$&$8.6(9)$&$$&$$&$22.0(55)$&$45.9(73)$&$8.6$\\
 & MR&$1.45$ & $1.63$&$1.76$&$1.88$&$16.94$ & $29.69$&$80.63$&$89.27$&$128.55$&$6.54$&0.0267\\
   &M&1.47 &1.70& 1.90 &2.11 &23.3& 45.7 &148 &84.6 &123.7& 6.5&0.0252\\
  &D&1.48 &1.72 &1.93 &2.14 &29.6 &57.5 &180 &82.9 &121.6 &6.4&0.0248\\
  &K&1.46 &1.68 &1.86 &2.06 &26.3 &48.2 &142 &87.6 &127.2 &6.6&0.0265\\\\

  $^{170}$Yb&Exp &$$&$$&$1.79(16)$&$1.77(14)$&$5.4(10)$&$$&$$&$13.4(34)$&$23.9(57)$&$2.4(6)$\\
 & MR&$1.46$ & $1.66$&$1.83$&$1.99$&$19.77$ & $36.95$&$111.68$&$61.71$&$89.74$&$4.66$&0.0395\\
  &M&1.44 &1.64 &1.78& 1.93 &8.3& 15.1 &44 &73.6 &106.3 &5.4&0.0315\\
  &D&1.47 &1.71 &1.91& 2.12& 30.6& 58.2 &176 &66.2 &97.1 &5.1&0.0633\\
  &K&1.47 &1.69 &1.90 &2.13& 31.9 &58.0& 168 &69.4 &101.3 &5.3&0.0646\\\\

 $^{172}$Yb&Exp &$1.42(10)$&$1.51(14)$&$1.89(19)$&$1.77(11)$&$1.1(1)$&$3.7(6)$&$12(1)$&$6.3(6)$&$$&$0.6(1)$\\
 & MR&$1.46$ & $1.67$&$1.85$&$2.03$&$20.40$ & $38.78$&$120.49$&$44.00$&$64.44$&$3.40$&0.0369\\
  &M&1.46 &1.67 &1.83 &1.99 &23.6 &43.1& 124 &48.7 &70.9 &3.7&0.0342\\
  &D&1.46 &1.67 &1.83 &1.99 &32.2 &55.9 &147 &51.6 &75.0 &3.9&0.0355\\
  &K&1.46 &1.66 &1.83 &2.01& 29.6 &51.6 &139 &51.0 &74.2 &3.8&0.0361\\\\

$^{174}$Yb&Exp &$1.39(7)$&$1.84(26)$&$1.93(12)$&$1.67(12)$&$$&$$&$$&$$&$12.4(29)$&$$\\
 & MR&$1.45$ & $1.63$&$1.77$&$1.90$&$17.43$ & $30.84$&$85.17$&$42.72$&$61.86$&$3.19$&0.0718\\
  &M&1.45 &1.63& 1.75& 1.86 &18.9 &32.3& 88& 44.6& 64.3& 3.3&0.0690\\
  &D&1.45& 1.63 &1.75 &1.86 &20.9 &35.1 &88 &45.0 &64.9& 3.3&0.0691\\
  &K&1.44 &1.62 &1.74 &1.86 &20.6 &34.3 &85 &45.7 &65.8 &3.4&0.0711\\\\

$^{176}$Yb&Exp &$1.49(15)$&$1.63(14)$&$1.65(28)$&$1.76(18)$&$$&$$&$$&$9.8$&$$&$$\\
 & MR&$1.45$ & $1.64$&$1.78$&$1.91$&$17.89$ & $31.96$&$89.72$&$62.34$&$90.11$&$4.63$&0.0413\\
  &M&1.44 &1.62 &1.74 & 1.87& 0.5 &1.0 &3 65.1& 93.7& 4.8&0.0317\\
  &D&1.46 & 1.66 &1.82 &1.97 &27.9& 49.0 &132& 63.1& 91.6& 4.7&0.0551\\
  &K&1.45 &1.65 &1.81 &1.97 &28.6 &49.0 &128 &64.5& 93.4& 4.8&0.0541\\\\

$^{174}$Hf&Exp &$$&$$&$$&$$&$14(4)$&$$&$9(3)$&$31.6(161)$&$48.7(124)$&$$\\
 & MR&$1.45$ & $1.65$&$1.81$&$1.96$&$19.10$ & $35.09$&$103.15$&$68.71$&$99.59$&$5.14$&0.0283\\
  &M&1.46 &1.66& 1.81 &1.97& 20.3 &37.3 &109 &65.4& 94.9& 4.9&0.0288\\
  &D&1.48 &1.74& 1.96 &2.20 &31.4 &62.2 &200 &66.9 &98.8 &5.3&0.0503\\
  &K&1.49 &1.76 &2.05 &2.42& 47.1 &87.5 &264 &69.7 &102.9& 5.5&0.0663\\\\

$^{176}$Hf&Exp &$$&$$&$$&$$&$5.4(11)$&$$&$31(6)$&$21.3(26)$&$$&$$\\
 & MR&$1.46$ & $1.67$&$1.85$&$2.04$&$20.55$ & $39.25$&$122.82$&$54.41$&$79.50$&$4.17$&0.0327\\
  &M&1.46 &1.68 &1.86& 2.04& 23.3& 43.9& 133& 57.0& 83.2 &4.4&0.0363\\
  &D&1.47 &1.70 &1.89 &2.09 &30.8 &57.3 &169 &57.9 &84.9 &4.5&0.0481\\
  &K&1.46 &1.68 &1.86& 2.06 &29.1 &52.2 &148& 60.3 &87.8 &4.6&0.0417\\\\

$^{178}$Hf&Exp &$$&$1.38(9)$&$1.49(6)$&$1.62(7)$&$0.4(2)$&$$&$2.4(9)$&$24.5(39)$&$27.7(28)$&$1.6(2)$\\
 & MR&$1.46$ & $1.67$&$1.85$&$2.03$&$20.41$ & $38.80$&$120.62$&$72.99$&$106.23$&$5.53$&0,0780\\
 &M&1.46 &1.68& 1.86 &2.04 &23.3 &44.3 &136 &73.5 &107.2& 5.6&0,0805\\
  &D&1.47& 1.69 &1.88 &2.07 &28.4& 53.1 &158& 73.8 &107.8 &5.6&0,0855\\
  &K&1.46 &1.68 &1.86 &2.06 &27.1 &49.2 &142 &75.7 &110.0& 5.7&0,0824\\\\

$^{180}$Hf &Exp&$1.48(20)$&$1.41(15)$&$1.61(26)$&$1.55(10)$&$$&$$&$$&$24.5(47)$&$32.9(56)$&$$\\
 &MR& $1.47$ & $1.70$&$1.92$&$2.16$&$22.20$ & $44.81$&$152.59$&$66.88$&$98.36$&$5.23$&0.1241\\
  &D&1.46 &1.66 &1.82 &1.98 &34.9& 59.5& 151& 78.4& 113.4 &5.8&0.0911\\
  &K&1.46 &1.66 &1.83 &2.00 &34.6 &58.6 &150 &80.3 &116.0 &6.0&0.0944\\\\

 $^{182}$W&Exp &$1.43(8)$&$1.46(16)$&$1.53(14)$&$1.48(14)$&$6.6(6)$&$4.6(6)$&$13(1)$&$24.8(12)$&$24.8(12)$&$0.2$\\
 & MR&$1.46$ & $1.66$&$1.83$&$2.00$&$19.78$ & $36.97$&$111.79$&$76.71$&$111.32$&$5.76$&0.0647\\
 &M&1.46 &1.68 &1.85& 2.00& 25.7 &47.4 &137 &79.3 &115.4& 6.0&0.0669\\
  &D&1.47& 1.69 &1.87& 2.04& 32.5& 58.3 &162 &79.9 &116.2 &6.0&0.0719\\
  &K&1.46 &1.67 &1.85 &2.05 &33.0& 57.5 &155 &82.0 &118.9 &6.1&0.0710\\\\

 $^{184}$W&Exp &$1.35(12)$&$1.54(9)$&$2.00(18)$&$2.45(51)$&$1.8(3)$&$$&$24(3)$&$37.1(28)$&$70.6(51)$&$4.0(4)$\\
 & MR&$1.49$ & $1.78$&$2.07$&$2.43$&$24.36$ & $55.19$&$220.78$&$108.02$&$160.65$&$8.71$&0.0401\\
  &M&1.48& 1.71& 1.88& 2.04& 30.5 &57.4& 167& 124.8& 182.0& 9.4&0.0575\\
  &D&1.48& 1.73 &1.95 &2.16 &40.7 &75.2& 216 &128.3 &187.3 &9.8&0.0494\\
  &K&1.48 &1.73 &1.97 &2.27 &38.9 &71.8 &214 &128.4 &187.1 &9.8&0.0420\\\\

$^{186}$W &Exp&$1.30(9)$&$1.69(12)$&$1.60(12)$&$1.36(36)$&$$&$$&$$&$41.7(92)$&$91.0(201)$&$$\\
 & MR&$1.39$ & $1.49$&$1.52$&$1.52$&$726.29$ & $1022.06$&$1780.46$&$168.58$&$240.37$&$11.94$&0.0569\\
 &M&1.49& 1.76 &1.99& 2.20 &31.4 &64.7& 213 &164.0 &241.2 &12.7&0.1604\\
  &D&1.51& 1.80 &2.07 &2.34 &46.2 &91.9& 289 &165.7 &244.5 &12.9&0.1874\\
  &K&1.49& 1.77 &2.08 &2.48 &47.3 &89.1 &275 &174.0 &254.4 &13.3&0.2081\\\\

$^{186}$Os &Exp&$1.45(7)$&$1.99(7)$&$1.89(11)$&$2.06(44)$&$$&$$&$$&$109.4(71)$&$254.6(150)$&$13.0(47)$\\
 & MR&$1.49$ & $1.77$&$2.05$&$2.40$&$24.18$ & $54.05$&$212.38$&$141.94$&$209.52$&$11.18$&0.0629\\
 &M&1.50& 1.80& 2.11 &2.45 &26.2 &59.5& 235 &163.5& 242.2 &13.0&0.0700\\
  &D&1.53& 1.87& 2.20 &2.55 &39.7 &90.2 &335 &164.9 &247.4& 13.4&0.0852\\
  &K&1.50& 1.81& 2.16 &2.63 &39.2 &81.0 &288 &173.5 &255.9 &13.6&0.0940\\\\

$^{188}$Os &Exp&$1.68(11)$&$1.75(11)$&$2.04(15)$&$2.38(32)$&$$&$$&$$&$63.3(92)$&$202.5(304)$&$43.0(74)$\\
 &MR& $1.51$ & $1.84$&$2.22$&$2.69$&$25.17$ & $62.78$&$283.13$&$228.55$&$339.59$&$18.25$&0.0654\\
  &M&1.51 &1.84 &2.22 &2.73 &23.0& 56.7& 257& 245.3& 363.5& 19.4&0.0713\\
  &D&1.54 &1.89 &2.25 &2.63 &33.9 &83.9 &344 &229.8 &345.2 &18.7&0.0626\\
  &K&1.52& 1.87 &2.29 &2.87 &33.6 &78.5 &330 &246.6 &366.2 &19.7&0.0903\\\\

 $^{230}$Th&Exp &$1.36(8)$&$$&$$&$$&$5.7(26)$&$$&$20(11)$&$15.6(59)$&$28.1(100)$&$1.8(11)$\\
 & MR&$1.45$ & $1.64$&$1.79$&$1.93$&$18.38$ & $33.20$&$94.91$&$64.73$&$93.67$&$4.82$&0.0231\\
  &M&1.47& 1.69 &1.88 &2.07& 23.6& 45.3& 141& 62.5& 91.4& 4.8&0.0296\\
  &D&1.47& 1.70 &1.90 &2.09 &30.0 &56.4 &168 &63.6 &93.2 &4.9&0.0331\\
  &K&1.46 &1.68 &1.86& 2.07& 31.4 &55.6 &155 &66.9 &97.3 &5.1&0.0311\\\\

$^{232}$Th&Exp &$1.44(15)$&$1.65(14)$&$1.73(12)$&$1.82(15)$&$14(6)$&$2.6(13)$&$17(8)$&$14.6(28)$&$36.4(56)$&$0.7$\\
 & MR&$1.45$ & $1.63$&$1.76$&$1.88$&$16.73$ & $29.21$&$78.75$&$60.22$&$86.84$&$4.43$&0.0118\\
  &M&1.46& 1.66& 1.82& 1.98& 21.3 &39.1& 115& 56.7 &82.4& 4.3&0.0219\\
  &D&1.46 &1.67& 1.84 &2.01& 25.8 &47.1 &135& 57.0 &83.0& 4.3&0.0261\\
  &K&1.45 &1.65 &1.80 &1.96 &26.0& 44.9 &119 &61.9 &89.6 &4.6&0.0203\\\\

$^{234}$U &Exp&$$&$$&$$&$$&$$&$$&$$&$12.5(27)$&$21.1(44)$&$1.2(3)$\\
 &MR& $1.44$ & $1.62$&$1.74$&$1.85$&$15.82$ & $27.15$&$71.03$&$43.90$&$63.30$&$3.23$&0.0175\\
 &M&1.45& 1.63 &1.77& 1.89& 18.5 &32.4 &88 &42.8 &62.0& 3.2&0.0170\\
  &D&1.45 &1.64& 1.78 &1.90& 20.7& 36.1& 97& 42.7& 61.8 &3.2&0.0169\\
  &K&1.45 &1.65 &1.80 &1.96 &26.0& 44.9 &119 &61.9 &89.6 &4.6&0.0282\\\\

$^{236}$U &Exp&$1.42(11)$&$1.55(11)$&$1.59(17)$&$1.46(17)$&$$&$$&$$&$$&$$&$$\\
 & MR&$1.44$ & $1.61$&$1.73$&$1.83$&$14.90$ & $25.17$&$64.00$&$45.34$&$65.25$&$3.32$&0.0996\\
 &M&1.45 &1.63 &1.75 &1.87 &17.6& 30.4 &80 &44.6 &64.4 &3.3&0.1115\\
  &D&1.45& 1.63 &1.76 &1.87 &19.3& 33.2 &87 &44.7 &64.5& 3.3&0.1124\\
  &K&1.44 &1.62 &1.75 &1.86 &19.5& 32.7 &82 &45.5 &65.6 &3.3&0.1086\\\\

$^{238}$U &Exp&$$&$$&$1.45(23)$&$1.71(22)$&$1.4(6)$&$3.6(14)$&$12(5)$&$10.8(8)$&$18.9(17)$&$1.2(1)$\\
 &MR& $1.44$ & $1.62$&$1.73$&$1.83$&$15.23$ & $25.88$&$66.47$&$38.61$&$55.65$&$2.84$&0.0389\\
 &M&1.45 &1.62& 1.75 &1.86 &17.2& 29.6 &77 &37.7 &54.4& 2.8&0.0429\\
  &D&1.45& 1.63 &1.75& 1.86& 18.9 &32.3 &83 &37.7& 54.5& 2.8&0.0430\\
  &K&1.44 &1.62 &1.74 &1.85 &19.3 &32.3 &80 &39.4 &56.8 &2.9&0.0414\\\\

$^{238}$Pu&Exp &$$&$$&$$&$$&$14(4)$&$$&$11(4)$&$$&$$&$$\\
 &MR& $1.44$ & $1.61$&$1.73$&$1.83$&$14.88$ & $25.13$&$63.88$&$41.24$&$59.38$&$3.02$&0.0262\\
 &M&1.44 &1.60 &1.70 &1.78 &6.1 &10.0 &24 &42.4 &60.8 &3.1&0.0075\\
  &D&1.44 &1.62 &1.73 &1.84 &19.1 &31.7 &78 &41.6 &59.9& 3.0&0.0334\\
  &K&1.44 &1.61 &1.73 &1.82 &18.8 &30.8 &74 &42.2 &60.7 &3.1&0.0314\\\\

$^{250}$Cf &Exp&$$&$$&$$&$$&$$&$$&$$&$6.8(17)$&$10.9(25)$&$0.6(1)$\\
 & MR&$1.44$ & $1.61$&$1.72$&$1.81$&$14.13$ & $23.58$&$58.63$&$39.83$&$57.29$&$2.91$&0.0190\\
 &M&1.44 &1.60& 1.71 &1.80 &13.0& 21.6 &53 &40.1& 57.7 &2.9&0.0192\\
  &D&1.44 &1.61& 1.72 &1.81 &15.0& 24.9& 61 &40.0& 57.5 &2.9&0.0191\\
  \hline
\end{longtable}
}

\clearpage

\end{document}